\documentclass[9pt,journal]{IEEEtran}

\usepackage{amsmath}

\usepackage[style=ieee]{biblatex}
\usepackage{pdfpages}

\begin{document}
\title{Tactile Perception of Electroadhesion: Effect of DC versus AC Stimulation and Finger Moisture}

\author{%%%% Author details
Easa AliAbbasi$^1$, Muhammad Muzammil$^1$, Omer Sirin$^1$, Philippe Lefevre$^{2,3}$, Ørjan Grøttem Martinsen$^{4,5}$, and Cagatay Basdogan$^{1,*}$}

\author{Easa~AliAbbasi, Muhammad~Muzammil, Omer~Sirin, Philippe~Lefèvre, Ørjan~Grøttem~Martinsen, and~Cagatay~Basdogan% <-this % stops a space
\IEEEcompsocitemizethanks{\IEEEcompsocthanksitem E. AliAbbasi, M. Muzammil, O. Sirin, and C. Basdogan are with the College of Engineering, Koc University, Istanbul 34450, Turkey (E-mail: ealiabbasi20@ku.edu.tr; mmuzammil22@ku.edu.tr;osirin13@ku.edu.tr; cbasdogan@ku.edu.tr).

P. Lefèvre is with Institute of Information and Communication Technologies, Electronics and Applied Mathematics (ICTEAM), Université catholique de Louvain, Brussels and Louvain-la-Neuve, Belgium and Institute of Neuroscience, Université catholique de Louvain, Brussels and Louvain-la-Neuve, Belgium (E-mail: philippe.lefevre@uclouvain.be).

Ø.G. Martinsen is with Department of Physics, University of Oslo, Sem Sælands vei 24, 0371 Oslo, Norway and Department of Clinical and Biomedical Engineering, Oslo University Hospital, 0424 Oslo, Norway (E-mail: o.g.martinsen@fys.uio.no).

 }
\thanks{$^{\#}$Corresponding author: \tt\small cbasdogan@ku.edu.tr}}
%\thanks{Manuscript received XX XX, 202X; revised XX XX, 202x.}}

%\markboth{IEEE TRANSACTIONS ON HAPTICS}{}%

\maketitle

\begin{abstract}
\textbf Electroadhesion has emerged as a viable technique for displaying tactile feedback on touch surfaces, particularly capacitive touchscreens found in smartphones and tablets. This involves applying a voltage signal to the conductive layer of the touchscreen to generate tactile sensations on the fingerpads of users. In our investigation, we explore the tactile perception of electroadhesion under DC and AC stimulations. Our tactile perception experiments with 10 participants demonstrate a significantly lower voltage detection threshold for AC signals compared to their DC counterparts. This discrepancy is elucidated by the underlying electro-mechanical interactions between the finger and the voltage-induced touchscreen and considering the response of mechanoreceptors in the fingerpad to electrostatic forces generated by electroadhesion. Additionally, our study highlights the impact of moisture on electroadhesive tactile perception. Participants with moist fingers exhibited markedly higher threshold levels. Our electrical impedance measurements show a substantial reduction in impedance magnitude when sweat is present at the finger-touchscreen interface, indicating increased conductivity. These findings not only contribute to our understanding of tactile perception under electroadhesion but also shed light on the underlying physics. In this regard, the results of this study extend beyond mobile devices to encompass other applications of this technology, including robotics, automation, space missions, and textiles. 
\end{abstract}


\IEEEpeerreviewmaketitle

\section{Introduction}
\IEEEPARstart{T}{he} human haptic sense is a remarkable sensory system capable of detecting nano-scale wrinkles on seemingly smooth surfaces \cite{skedung2013feeling} and distinguishing between smooth surfaces with different material coatings \cite{aliabbasi2023tactile} or even modified surface chemistries \cite{carpenter2018human}. Despite the extraordinary capabilities of the human finger in discerning minute details, there remains a limited number of actuation technologies that can artificially replicate similar tactile sensations on touch surfaces. Surface haptics, an emerging field of research, aims to improve the way users interact with touch surfaces such as the touchscreens of mobile devices, enhancing the user experience by providing realistic and finer tactile feedback \cite{basdogan2020review}. In this regard, electroadhesion (EA) via electrostatic actuation appears to be a promising technique for displaying frictional forces to the user's finger as it moves on the touchscreen. In this technique, a voltage signal is applied to the conductive layer of a capacitive touchscreen to generate an electrostatic attraction force between its surface and the finger sliding on it \cite{bau2010teslatouch,meyer2013fingertip,sirin2019electroadhesion,aliabbasi2022frequency}. This results in an increase in the frictional force acting against the finger, in the direction opposite to its movement. Although the technology for generating tactile feedback on a touchscreen via EA is already in place and straightforward to implement, our knowledge of the underlying contact mechanics, the nature of electrical interactions between the human finger and the touchscreen, and also our perception of tactile stimuli generated by EA are still limited. Unraveling the physics behind EA holds the promise of unlocking innovative technological applications. Beyond mobile devices, where the potential includes experiencing digital shapes and textures on touch surfaces \cite{osgouei2017improving,vardar2017roughness,icsleyen2019tactile,friesen2021building,sadia2022exploration} and interacting with them through finger/hand gestures \cite{nakamura2016multi, emgin2018haptable}, these advancements are poised to extend into diverse domains such as robotics, automation, space missions, and textiles (see a more comprehensive review of EA applications in \cite{guo2019electroadhesion, rajagopalan2022advancement}).

In terms of contact mechanics, only a few studies have recently shed some light on the physics behind EA. The change in electrostatic forces between the human finger and a voltage-induced touchscreen is observed to be proportional to the square of the voltage amplitude \cite{meyer2013fingertip}. The increase in electrostatic force due to EA results in a 0.25\% increase per Volt in frictional force compared to the case of no EA (e.g. 25\% increase for 100 V) \cite{basdogan2020modeling}. It was claimed that the increase in frictional force is due to an increase in the real contact area \cite{ayyildiz2018contact}. Despite the observed decrease in the measured apparent contact area during sliding under EA \cite{sirin2019fingerpad}, the rise in the number of microscopic contacts at the interface due to EA leads to an increase in the real contact area. This hypothesis aligns well with the contact mechanics theory proposed for EA by Persson \cite{persson2018dependency,persson2021general}, which considers the multi-scale nature of contacting surfaces.

Compared to the studies on contact mechanics, the number of studies investigating the electrical interactions between a human finger and a touch surface under EA is only a few. Earlier studies \cite{shultz2018electrical,vardar2021finger, aliabbasi2024experimental} showed that the electrical impedance of the interfacial gap between the finger and the touch surface is significantly lower for the stationary finger compared to that of the sliding finger. It was suggested that when the finger remains stationary on the touchscreen, sweat accumulates at the interfacial gap and reduces the potential difference, decreasing the magnitude of the electrostatic forces \cite{shultz2018electrical}. Our recent study \cite{aliabbasi2022frequency} highlighted the important role of charge leakage from the Stratum Corneum (SC), the outermost layer of skin, to the touch surface at low frequencies. The experimental results showed that the electrostatic force exhibits an inverted parabolic curve with a peak value at around 250 Hz. An electromechanical model based on the fundamental laws of electric fields and Persson's contact mechanics theory \cite{persson2001theory} was developed to estimate the frequency-dependent magnitude of electrostatic forces. The model revealed that the electrical properties of the SC and the charge leakage from it are the main causes of the inverted parabolic behavior.

In terms of tactile perception of EA, the number of studies is also limited. The sensitivity of the human finger to the polarity of the voltage signal was investigated and the results showed that the participants perceived negative or biphasic pulses better than positive ones \cite{kaczmarek2006polarity}. The detection and discrimination threshold voltages across different frequencies were measured through psychophysical experiments, revealing a statistically significant relationship between absolute detection voltage and signal frequency. The results showed a U-shaped curve in detection threshold voltage, with the lowest value at 125Hz whereas the discrimination voltage remained constant at 1.16 dB against all tested frequencies \cite{bau2010teslatouch}. The variation in tactile perception corresponding to the changing waveform of the applied voltage was investigated in \cite{vardar2017effect}. The results showed that the participants were more sensitive to square voltage signals than sinusoidal ones for frequencies lower than 60 Hz. The analysis of the collected force and acceleration data in the frequency domain, by considering the human tactile sensitivity curve, suggested that the Pacinian channel was predominantly responsible for detecting EA stimuli. This was consistent across all square wave signals displayed at various frequencies. The interference of multiple tactile stimuli (tactile masking) under EA was also investigated \cite{vardar2018tactile}. The results showed that the sharpness perception of virtual edges depends on the masking amplitude and activation levels of frequency-dependent psychophysical channels. The tactile perception of a step change in friction due to EA was investigated by considering the influence of normal force and sliding velocity \cite{ozdamar2020step}. Participants perceived rising friction (EA is switched from OFF to ON during sliding) as stronger than falling friction (EA is switched from ON to OFF during sliding), and both the normal force and sliding velocity significantly influenced their perception. %Increasing sliding velocity increased the perception whereas increasing normal force reduced the perception.

With the current technology, generating electrostatic forces between a human finger and a touchscreen to modulate the frictional forces between them is quite straightforward. Empirical evidence indicates that generating a perceptible tactile sensation requires the utilization of an AC voltage signal. A DC input voltage signal, even with a higher amplitude, does not appear to generate a similar sensation, but the root causes of this difference are not fully known yet. Moreover, finger moisture and environmental humidity are known to affect frictional forces under EA \cite{sirin2019fingerpad}, but their effect on tactile perception and the physics behind the change in electrostatic force intensity due to accumulated sweat at the interface has received less attention. One study investigated the effect of electrowetting, the change in the wettability of the liquid on the touchscreen under EA, and concluded that the increase in frictional forces between the finger and the touchscreen at higher humidity levels is mainly due to the increase in capillary forces \cite{li2020electrowetting}. It was also observed that the finger left more residue (primarily, sweat and sebum) in the areas of a touchscreen where EA is active \cite{chatterjee2023preferential}. The authors suggested that i) the electrohydrodynamic deformation of sebum droplets adhere to the finger valleys, which results in the creation of extra capillary bridges and leftover droplets on the screen's surface after they break, and ii) the electric field-induced stabilization of sebum capillary bridges exist between the finger ridges and the screen, which leads to the merging and formation of larger droplets.

In this study, we investigate the differences in human tactile detection threshold under EA for the DC and AC voltage signals applied to the touchscreen. We show that the detection threshold under the AC stimulation is significantly lower than that of the corresponding DC stimulation. We argue that both perceptual and physical mechanisms cause this discrepancy. We explain the perceptual mechanism behind this discrepancy using the tactile sensitivity of the human finger to frequency-dependent stimulation of electrostatic forces under EA \cite{vardar2017effect} while an electrical circuit model, developed based on our electrical impedance measurements, is utilized to explain the physics behind it. 

We also highlight the adverse effect of fingertip skin moisture on the tactile perception of EA by correlating the finger moisture of the participants measured by a corneometer with their threshold values for the AC signal. We show that the magnitude of electrostatic forces inferred from friction measurements is lower for a wet finger than for a dry finger. Our electrical impedance measurements suggest that the impedance drops drastically when there is liquid at the interface. As a result, the voltage difference at the air gap and the magnitude of electrostatic forces decrease.

\section{Material and Methods}\label{sec:experimental section}
\subsection{Participants}
The tactile perception experiment was conducted with ten adult participants (four females and six males) having an average age of 29.4 years (SD: 5.9). Due to its time-consuming nature, the electrical impedance and friction measurements were performed with one male participant (age: 34 years old) having relatively dry fingers. All participants provided written informed consent to undergo the procedure, which was approved by the Ethical Committee for Human Participants of Koc University (Protocol Numbers: 2022.128.IRB2.020, 2023.280.IRB2.060). The investigation conformed to the principles of the Declaration of Helsinki and the experiments were performed by following relevant guidelines and regulations.

\begin{figure*}[!t]
\centering
  \includegraphics[width=2\columnwidth]{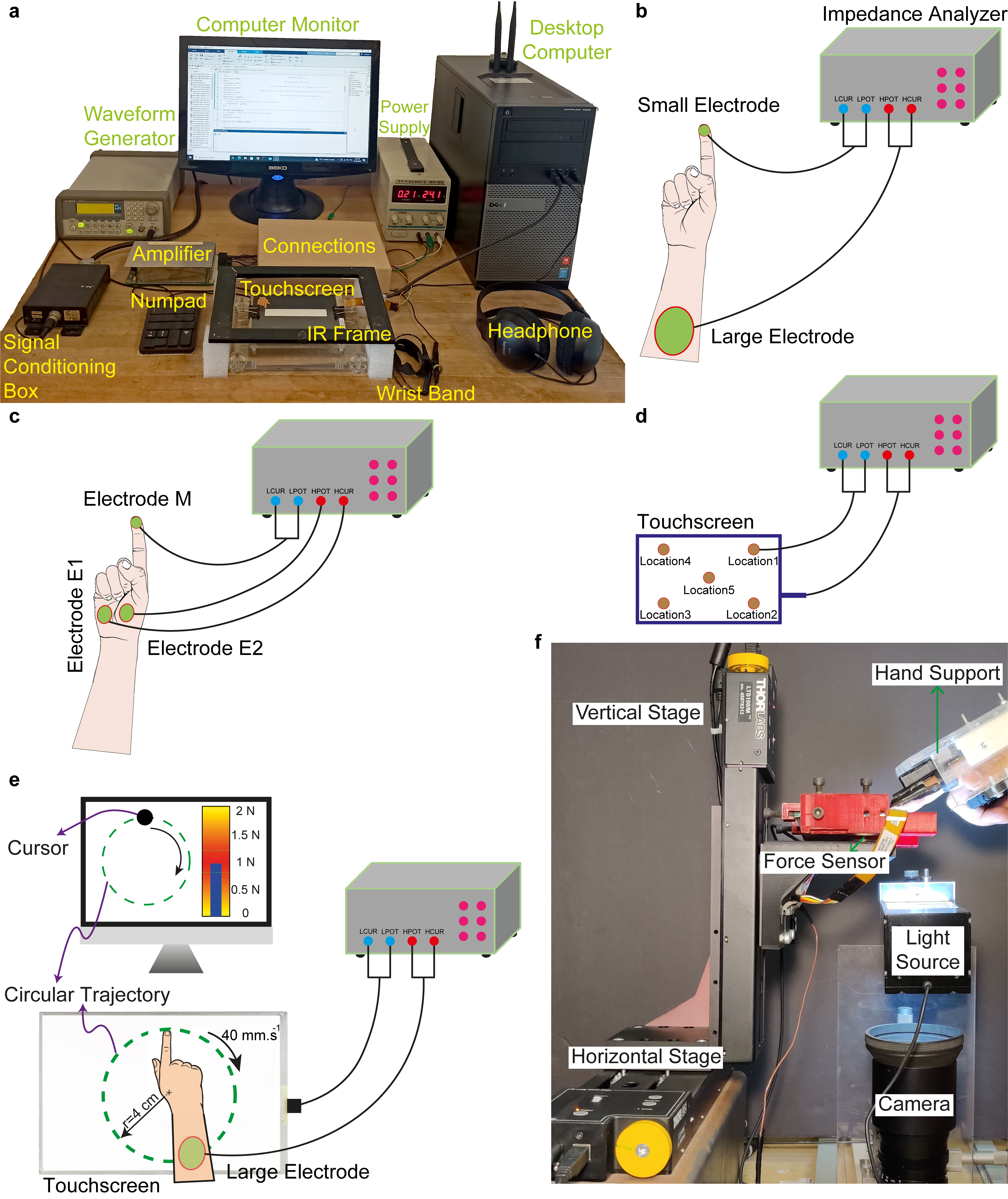}
  \caption{a) The setup used in our tactile perception experiments. Schematic representation of b) electrical impedance measurement for skin, c) skin moisture level assessment, d) electrical impedance measurement for touchscreen, and e) electrical impedance measurement for the finger sliding on the touchscreen. f) The setup used in our friction force measurements.}
  \label{fig:Setup}
\end{figure*}

\subsection{Tactile Threshold Experiments}
We used the setup shown in Fig. \ref{fig:Setup}a for the tactile perception experiment. In this setup, the input voltage signals were generated by a waveform generator (33220A, Agilent Inc.) connected to a PC via a TCP/IP protocol. The signals were then amplified by a piezo amplifier (E-413, PI Inc.) and applied to the conductive layer of a capacitive touchscreen (SCT3250, 3M Inc.) for displaying tactile stimulus to the participants. The touchscreen was rigidly fixed with holders to avoid undesired mechanical vibrations during the experiments. A DC power supply (MCH-303D, Technic Inc.) was utilized to provide 24 V DC voltage for operating the amplifier. A high-resolution force sensor (Nano 17-SI-12-0.12, ATI Industrial Automation Inc.) was placed below the touchscreen to measure contact forces. These forces were acquired by a 16-bit analog data acquisition card (PCI-6034E, National Instruments Inc.) with 10 kHz sampling frequency. An IR frame (IRTOUCH Inc.) was placed above the touchscreen to detect finger position.

Before the experiments, the participants washed their hands with soap, rinsed with water, and dried them at room temperature, and the touchscreen was cleaned with alcohol. Throughout the experiments, the participants were asked to wear an elastic strap on their stationary wrist for grounding and put on headphones playing white noise to prevent their tactile perception from being affected by any external auditory cue.

To investigate human tactile detection thresholds for DC and AC voltage stimulations, we conducted the absolute detection experiment using 2-Alternative Forced-Choice (2AFC) paradigm \cite{vardar2016effect}. During the experiments, participants were asked to slide their index fingers on the touchscreen from left to right twice for a distance of 100 mm in each trial. The tactile stimulus was displayed only in one of the passes, which was randomized to eliminate any bias. Participants were asked to determine the pass (interval) in which they felt a tactile effect. To regulate their scan speed, a visual cursor moving at a speed of 20 mm/s was displayed on the computer monitor and the participants were asked to follow it with their index finger. To assist the participants with controlling their applied normal forces on the touchscreen, another visual feedback displayed the real-time magnitude of the applied normal force. The normal force and scanning speed were recorded in each trial. The average normal force for all participants was 0.334 N (SD: 0.084) and 0.332 N (SD: 0.073) under DC and AC stimulations, respectively. The average speed for all participants was 20.01 mm/s (SD: 1.34) and 20.06 mm/s (SD: 0.87) under DC and AC stimulations, respectively. If a participant's normal force or scan speed was not in the desired range (0.1-0.6 N; 10-30 mm/s), the trial was repeated until a measurement within the range was obtained. Before starting the experiments, participants were given instructions and asked to complete a training session. This training session enabled participants to adjust their finger scanning speed and normal force before the actual experimentation.

The amplitude of the voltage signal applied to the touchscreen (and hence the magnitude of the tactile stimulus) was altered using the three-down/one-up adaptive staircase method \cite{vardar2018tactile}. To determine the detection thresholds for the sinusoidal AC signal at 125 Hz and the corresponding DC signal, the experiment started with a voltage amplitude of 200 V$_{pp}$ and 70.7 V, respectively. It is worth noting here that the initial voltage amplitude provided sufficiently high intensity for all participants under the AC stimulation but not under the DC stimulation. However, for safety reasons, the maximum voltage was limited to 100 V in DC experiments. If a participant gave three correct responses (not necessarily consecutive), the voltage level was decreased by 5 dB. If a participant gave one incorrect response, the voltage level was increased by 5 dB. The change in response from correct to incorrect or vice versa was counted as one reversal. After one reversal, the step size was decreased to 1 dB. The experiments were stopped automatically if the reversal count was five at the ±1 dB level (Fig. S1a and b, Supplementary Material). The threshold was calculated as the mean of the last five reversals. The moisture level of each participant's skin was measured by a Corneometer (CM 825, Courage - Khazaka Electronic) four times just before and right after the experiment and an average of eight measurements was reported for each participant.

\subsection{Electrical Impedance Experiments}
We selected the four-electrode method for all our electrical impedance measurements performed by the impedance analyzer (MFIA, Zurich Instruments Inc.) except for the assessment of skin moisture level, in which we utilized three electrodes. Before each measurement session, the impedance analyzer was calibrated with the short-open-load option of the device to compensate for the residual impedances in the system.

\subsubsection{Electrical Impedance of Skin}
Fig. \ref{fig:Setup}b presents the schematic of the electrical impedance measurements for the skin. We measured the electrical impedances of the skin in three sessions on three consecutive days using hydrogel electrodes (1050NPSM Neonatal Pre Wired Small Cloth ECG Electrodes, Cardinal Health Inc.) attached to the participant's right-hand index finger. A larger electrode (HeartStart FR2 Defibrillator Electrode Pads, Philips Medical Systems Inc.) with a contact area approximately ten times larger than the hydrogel electrode was attached to the ventral forearm of the same hand. Using a custom-made circular tube, a weight of 100 grams was placed on top of the small electrode at the fingertip and kept vertically aligned. The weight was equivalent to a normal force of 1 N.

\subsubsection{Assessment of Skin Moisture Level}
The schematic representation of the skin moisture level assessment is shown in Fig. \ref{fig:Setup}c. We measured the low-frequency susceptance of the skin, an indicator of skin moisture \cite{martinsen1995electrical,martinsen1998using,martinsen2008gravimetric}, using the impedance analyzer (MFIA, Zurich Instruments Inc.). For all measurements, we selected the three-electrode measurement method, and the impedance analyzer was calibrated before every measurement session with the short-open-load option of the device to compensate for the residual impedances in the system. Electrode M in Fig. \ref{fig:Setup}c is a custom-made metal electrode attached to the participant's fingertip. Electrodes E1 and E2 are electrocardiogram (ECG) electrodes (Red Dot 2228, 3M Inc.), which were attached to the palm of the same hand. Using a custom-made circular tube, a weight of 100 grams was placed on top of Electrode M and kept vertically aligned. The weight was equivalent to a normal force of 1 N. The magnitude and phase of the electrical impedance were measured at 125 Hz stimulation frequency for 10 seconds, with a sampling frequency of 2.5 kHz. Before the experiment, the participant washed his hands with water and soap and dried them with a clean towel. In the nominal finger condition, the participant waited for five minutes in the experiment room, allowing his body to reach normal hydration levels. Conversely, in the moist finger condition, the participant wore thick clothing and waited for fifteen minutes in the experiment room to induce sweating and achieve an elevated level of skin moisture.

Once the magnitude and phase of the electrical impedance are known, one can calculate the real and imaginary parts of the impedance as:
\begin{equation*}
    Re\{Z\} = |Z|\cos{\Phi}
\end{equation*}
\begin{equation*}
    Im\{Z\} = |Z|\sin{\Phi}
\end{equation*}
where, $|Z|$ and $\Phi$ are the magnitude and phase of the measured electrical impedance, respectively. Hence, the susceptance can be calculated as:
\begin{equation*}
    B=-\frac{Im\{Z\}}{Re\{Z\}^2 + Im\{Z\}^2}
\end{equation*}

\subsubsection{Electrical Impedance of Touchscreen}
The schematic representation of the electrical impedance measurements for the touchscreen is presented in Fig. \ref{fig:Setup}d. Five metallic custom-made electrodes were manufactured for this specific measurement. Using a very thin layer of silver grease (8463A, MG Chemicals), the electrodes were attached to the touchscreen's surface at five different locations. These locations were selected to cover the surface of the touchscreen. The electrical impedance data were collected in ten consecutive trials from each location. Due to the sensitivity of the electrical impedances of the touchscreen to the thickness of the silver grease between the electrodes and the touchscreen, we tried to perform the measurements with the thinnest possible layer of silver grease. However, still relatively large variations are observed in the impedance measurements of the touchscreen.

\subsubsection{Total Electrical Impedance of Sliding Finger on Touchscreen}
Fig. \ref{fig:Setup}e presents the schematics for the total electrical impedance measurements while the participant's finger was sliding on the surface of the touchscreen. The measurements were performed under two different lubrication conditions: a) nominal and b) wet. In the nominal finger condition, no excess liquid was added to the interface between the finger and the touchscreen. However, obtaining consistent impedance data with a moist finger proved more challenging, attributed to variations in moisture level, stemming from the prolonged duration of the experimentation due to the frequency sweep procedure. For this reason, four droplets of 5 $\mu$L 0.9\% Isotonic Sodium Chloride (NaCl) were applied at four different locations on the touchscreen (a total of 20 $\mu$L) to imitate a moist finger (called 'wet' condition in the text). The measurements for each condition were performed in three separate sessions on three different days and the data were collected in ten consecutive trials. A user interface was developed to guide the participant in moving his finger on a circular path with a desired velocity while controlling his normal force. A black-colored cursor circled on a dashed-green trajectory ($r=4$ cm) with a constant velocity of 40 mm/s on the computer monitor and the participant was asked to follow its motion by moving his finger on the touchscreen at the same speed. The magnitude of the normal force applied by the participant's finger to the touchscreen was displayed by a blue bar on the monitor for visual feedback. The normal force was acquired by the force transducer (Nano 17-SI-12-0.12, ATI Industrial Automation Inc.) placed beneath the touchscreen. The participant was trained before the actual experiment to practice maintaining the normal force close to 1 N while his finger circled on the touchscreen.

\subsection{Friction Force Experiments}
Fig. \ref{fig:Setup}f shows the setup used for friction force measurements. It comprises a capacitive touchscreen (SCT3250, 3M Inc.), actuated by a voltage signal applied to its conductive layer, which was generated through a data acquisition card (PCIe-6321, National Instruments Inc.) and amplified by a piezo amplifier (PZD700A M/S, Trek Inc.). The setup was configured to keep the finger stationary while the touchscreen slid under it. The movements of the touchscreen were controlled by two translational stages (LTS150, Thorlabs Inc.) along the directions normal and tangential to the surface of the touchscreen. A high-speed camera (IL5H, Fastec Imaging Inc.) and a coaxial light source (C50C, Contrastech Inc.) were placed beneath the touchscreen to capture high-resolution images of the fingerpad contact area. A force transducer (Mini40-SI-80-4, ATI Industrial Automation Inc.) was attached under the touchscreen to measure the normal and tangential forces acting on the fingerpad via a data acquisition card (PCIe-6034E, National Instruments Inc.) at a sampling frequency of 1 kHz. A proportional-integral-derivative (PID) controller was implemented to maintain a constant normal force between the fingerpad and the touchscreen while the finger slides on its surface.

The experiment aimed to measure the friction force between the finger and the touchscreen under EA = OFF (no voltage was applied to the touchscreen) and EA = ON (an AC voltage signal of 120 V at a frequency of 125Hz was applied to the conductive layer of the touchscreen) for two distinct moisture conditions of the fingerpad: a) nominal and b) moist. The experiments for the nominal finger condition were conducted early in the morning from 7 am to 9 am, while the participant was fasting to minimize sweat generation. In contrast, the experiments for the moist finger condition were carried out in the afternoon between 2 pm to 5 pm. During this time, the participant wore thick warm clothes to raise body temperature, inducing increased sweat generation. It is imperative to note that no artificial liquid was introduced to the interface in either of the two experimental conditions. Before the experiments, the touchscreen was cleaned with alcohol. During the experiments, the index finger of the participant's right hand was placed in a custom-made hand support to ensure consistent contact with the touchscreen at an angle of 20 degrees. An electrical grounding strap was wrapped around the participant's wrist to keep him grounded when the voltage was applied to the touchscreen. The participant was advised to maintain a stable and stationary position throughout the experiments. The normal force applied to the touchscreen by the finger of the participant was maintained at 1 N via the PID controller in all trials. In each trial, the touchscreen was translated under the fingerpad of the participant in the tangential direction for a distance of 60 mm at a constant speed of 20 mm/s. 

There were a total of 108 trials in the experiment, performed in 3 days. Hence, there were 36 trials on each day (2 moisture conditions: nominal and moist $\times$ 2 EA states: OFF and ON $\times$ 3 trials$/$session $\times$ 3 sessions). For each trial, the CoF was calculated by dividing the recorded tangential force with the corresponding normal force. The electrostatic force acting on the finger was then inferred from the experimental CoF data \cite{basdogan2020modeling}:
\begin{equation}\label{eq:electrostatic Force}
    F_{e} = \left(1 - \frac{\mu^{OFF}}{\mu^{ON}}\right) F_{n}
\end{equation}
where $\mu^{ON}$ and $\mu^{OFF}$ represent measured CoF when EA = ON and EA = OFF, respectively.

\section{Results and Discussion}\label{sec:results and discussion}
\subsection{Tactile Threshold Measurements}
\begin{figure*}[!t]
  \includegraphics[width=2\columnwidth]{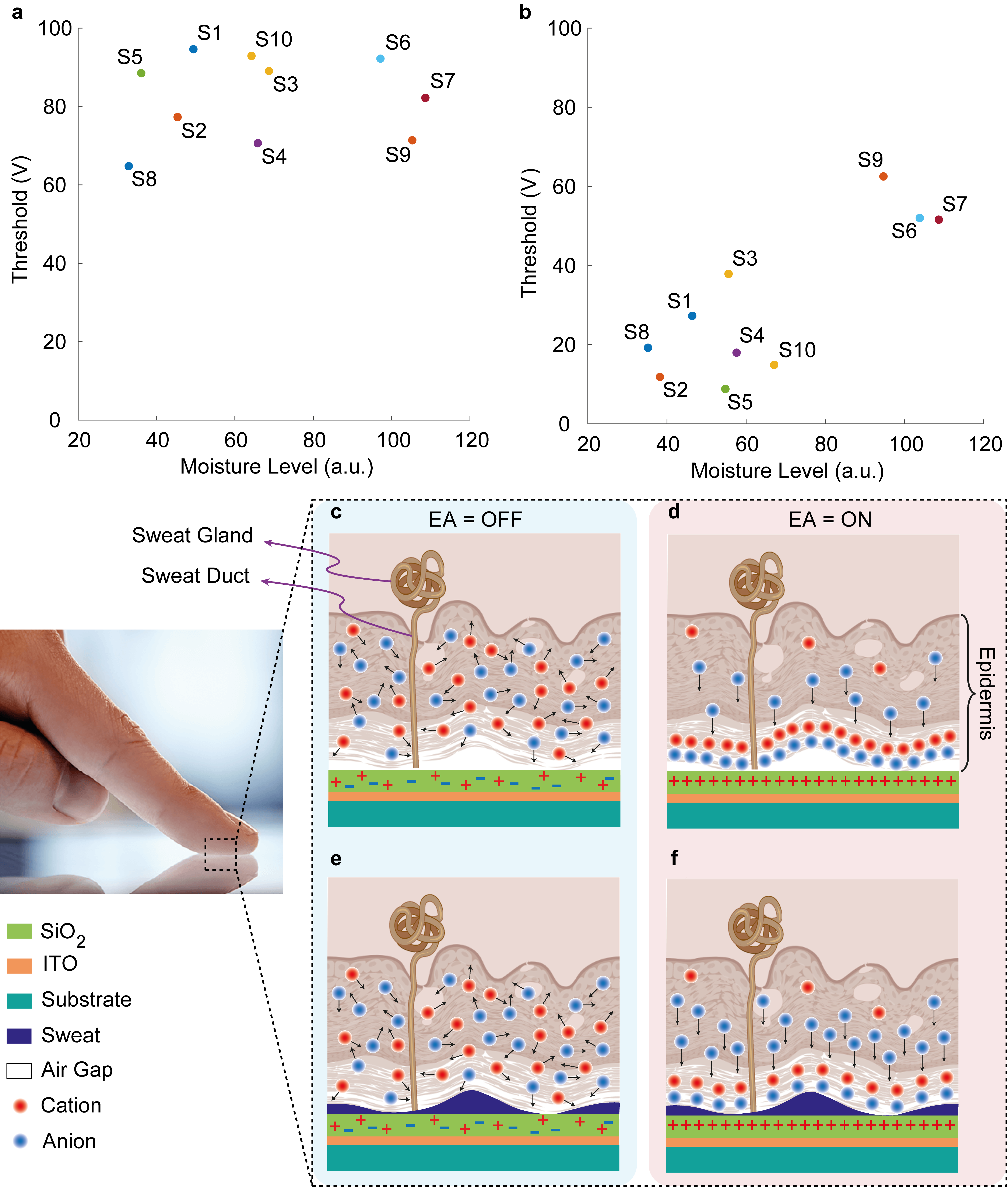}
  \caption{The threshold voltage versus moisture level of each participant under a) DC and b) AC stimulations. Moisture measurements are given in arbitrary units (a.u.) that range from 20 (dry skin) to 120 (very wet skin). c) Schematic representation of a human finger in contact with the surface of a touchscreen when EA = OFF, where the electrical charges and ions are distributed randomly. d) The schematic of a human finger in contact with the surface of the touchscreen under EA. Due to the formation of an electric double layer at the skin interface, electrical charges can leak to the surface of the touchscreen. Schematic representations of a human finger in contact with the surface of a touchscreen e) without EA and f) with EA when there is a layer of finger sweat at the interface. Due to its conductive nature, the charge exchange between the finger skin and the touchscreen increases.}
  \label{fig:Introduction}
\end{figure*}

Fig. \ref{fig:Introduction}a and b present the threshold voltages and fingertip skin moisture level of each participant under DC and AC stimulations, respectively. Exemplar data of the threshold experiment is given in Fig. S1a and b (Supplementary Material) for AC and DC stimulations, respectively. The results showed that the threshold voltage under AC stimulation (30.41 ± 19.28) was significantly lower than that of the DC stimulation (82.36 ± 10.72) for all participants. There was a clear convergence in threshold values for all participants under the AC stimulation, while there was no convergence under the DC stimulation (note that the experiments were stopped automatically if the reversal count was five at ±1 dB level). The threshold voltage was calculated as the mean of the last five reversals. The average threshold value obtained under AC stimulation (30.41 V) is consistent with the threshold value reported in our earlier study for the stimulation frequency of 125 Hz \cite{vardar2017effect}. We performed ANOVA with repeated measures on the moisture levels of participants using the measurement time (before and after the experiment) and the signal type (DC vs. AC) as the main factors. We did not observe a significant difference between the moisture levels before and after the experiments (F (1,9) = 0.205, p = 0.661) and DC versus AC (F (1,9) = 5.126, p = 0.051). However, the threshold voltages of the participants having very moist fingers (S6, S7, S9) were higher than the other participants under AC stimulation (see Fig. \ref{fig:Introduction}b). The Pearson correlation showed a positive and strong correlation between threshold voltages and moisture level (r = 0.81, p $<$ 0.01). This result is consistent with the results of our earlier study \cite{sirin2019fingerpad} and supports our claim that moisture has an adverse effect on the capacity of EA to modulate friction, which in turn affects the tactile perception of EA.

\subsubsection{Effect of Voltage Type}
Fig. \ref{fig:Introduction}c shows the schematic representation of a human finger in contact with the surface of a touchscreen when EA = OFF. The electrical charges in the insulator layer of the touchscreen (SiO$_2$) and ions in the SC layer of the skin distribute randomly. Under EA, the positive electrical charges in the insulator layer of the touchscreen attract the anions in the skin and accumulate at the skin's interface, creating the first layer of ions (see Fig. \ref{fig:Introduction}d). These anions attract the cations and build the second layer of ions on top of the first layer. This is known as the electric double layer. As the voltage difference between the touchscreen and the finger skin increases, the free anions in the skin get more attracted toward the interface and push the ions in the electric double layer as suggested in \cite{aliabbasi2024experimental}. Since ions cannot pass to the insulator layer of the touchscreen, their negative charges (electrons) jump to the surface of the touchscreen. The transfer of electrical charges from the finger skin to the touchscreen's surface is called the charge leakage phenomenon \cite{aliabbasi2022frequency}, which occurs at low stimulation frequencies. Hence, as the frequency approaches the DC stimulation, more leakage is observed, causing a reduction in the magnitude of the electric field and hence, the magnitude of electrostatic forces.

\subsubsection{Effect of Moisture}
Similar to the adverse effect of charge leakage at low frequencies, finger moisture produced by sweat glands also negatively affects the strength of EA. There are more than 500 eccrine sweat glands on human fingertips \cite{tripathi2015morphology} and the perspiration rate is fast enough to (partially) fill the air gap between the finger and the touchscreen, as shown in Fig. \ref{fig:Introduction}e. Under EA, the air gap between the finger and the touchscreen acts as an insulator and does not let the charges pass from the interface easily. However, when the air gap is replaced with sweat (Fig. \ref{fig:Introduction}f), charges can easily pass through the layers since sweat is a better conductor. Hence, the voltage at the air gap and the magnitude of electrostatic forces decrease.

\begin{figure*}[!t]
\centering
  \includegraphics[width=2\columnwidth]{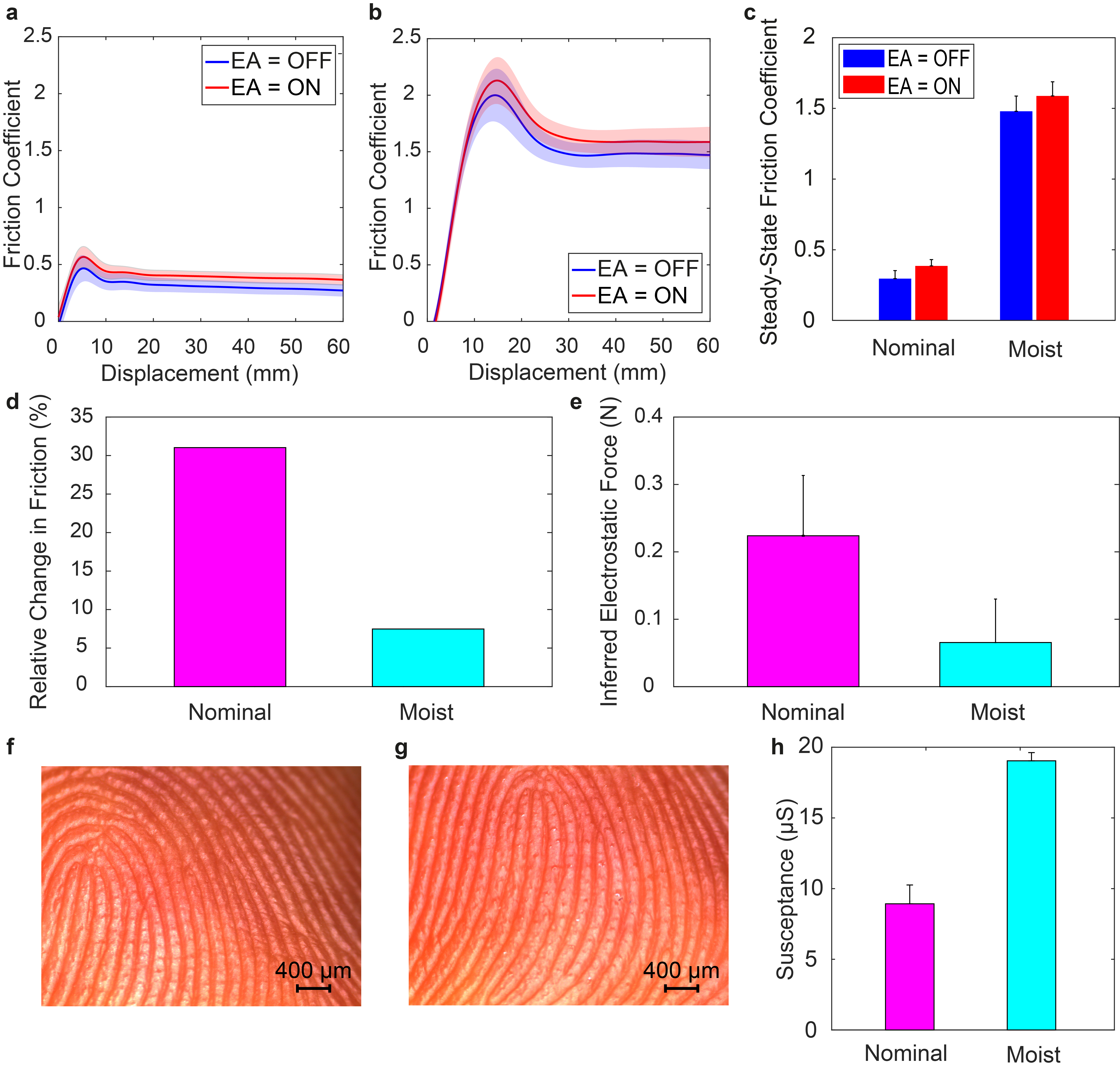}
  \caption{Coefficient of friction (CoF) as a function of displacement with and without EA for a) nominal and b) moist finger conditions. The solid curves show the mean values and the shaded regions around them are the standard deviations for all trials. c) Mean values of steady-state CoF under nominal and moist finger conditions with and without EA. d) Relative difference in CoF between EA = ON and EA = OFF for the nominal and moist finger conditions. e) Electrostatic forces inferred from the friction measurements for the nominal and moist finger conditions. f) An image captured from the participant's fingerpad in the nominal condition (5 minutes after resting in the experiment room). g) An image captured from the fingerpad of the same participant in the moist condition (15 minutes after resting in the experiment room while wearing thick clothing to activate the sweat glands): sweat accumulation is evident upon closer inspection of the image. h) Mean values of the susceptance, an indicator of skin moisture, measured under the nominal and moist conditions.}
  \label{fig:Friction}
\end{figure*}

\subsection{Friction Force Measurements}
 Fig. \ref{fig:Friction}a and b display the change in CoF as a function of displacement for the nominal and moist finger conditions respectively, where each curve represents the mean values of 27 trials. Blue and red-colored curves depict the CoF for EA = OFF and EA = ON, respectively. The shaded regions around the curves represent the standard deviations. The individual trials for the CoF under the nominal and moist conditions are presented in Fig. S2 and S3 (Supplementary Material) and Fig. S4 and S5 (Supplementary Material), respectively. Similarly, the individual trials recorded for the tangential forces under the nominal and moist conditions are reported in Fig. S6 and S7 (Supplementary Material) and Fig. S8 and S9 (Supplementary Material), respectively. 
 Fig. \ref{fig:Friction}c depicts the steady-state values of CoF computed by averaging the data within the interval between 35 mm to 45 mm of displacement. In the nominal finger condition, the mean values for EA = OFF and EA = ON were 0.29 ± 0.05 and 0.38 ± 0.04, respectively. For the moist condition, the mean values for EA = OFF and EA = ON were 1.47 ± 0.11 and 1.58 ± 0.10, respectively.

The relative differences in CoF between EA = ON and EA = OFF for the nominal and moist conditions are given in Fig. \ref{fig:Friction}d. There was a contrast of 31\% in the nominal condition, whereas a relative difference of around 7\% was observed in the moist condition. The electrostatic force under the nominal and moist conditions is given in Fig. \ref{fig:Friction}e. Fig. \ref{fig:Friction}f and g show the images taken from the participant's fingerpad under the nominal and moist conditions, respectively. Sweat started to come out of the sweat ducts under the moist condition as shown in Fig. \ref{fig:Friction}g. Fig. \ref{fig:Friction}h presents the mean values of fingerpad's susceptance under the nominal and moist conditions. The results showed that the susceptance of the moist finger was significantly higher than that of the nominal finger, suggesting that the moist finger had a relatively higher moisture level than that of the nominal finger, as anticipated. 

\begin{figure*}[!t]
\centering
  \includegraphics[width=2\columnwidth]{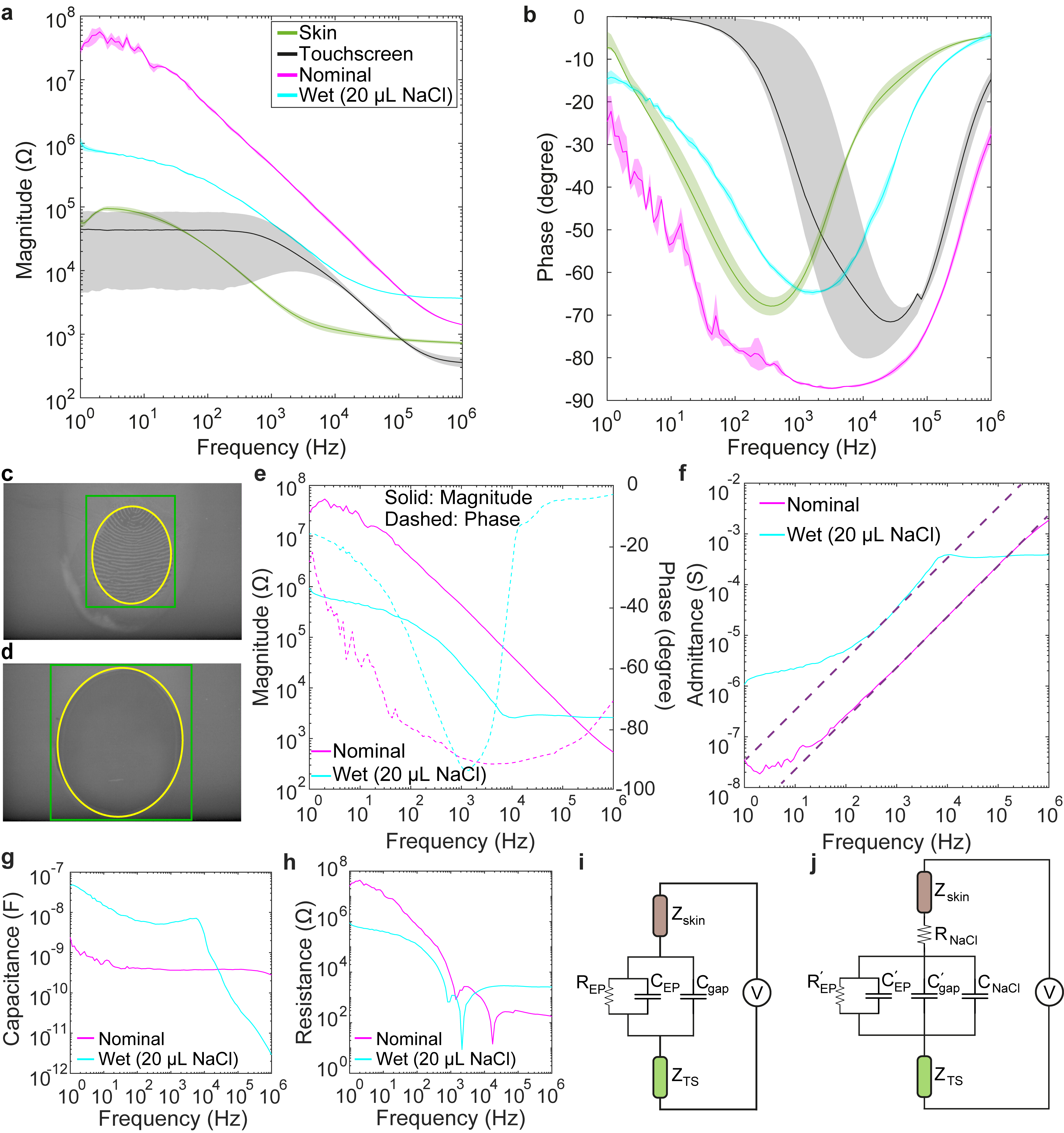}
  \caption{The change in average electrical impedance a) magnitude and b) phase as a function of frequency for finger skin (green), the touchscreen itself (black), sliding finger on the touchscreen under the nominal (magenta) and wet (cyan) conditions. Images of the participant’s fingerpad in contact with the surface of the touchscreen for a normal force of 1 N (no sliding): c) nominal and d) wet conditions. The green-colored rectangle and yellow-colored contour represent the region of interest for the image processing operations and fitted ellipse to the apparent contact area, respectively. e) The magnitude (solid curves) and phase (dashed curves) of the remaining impedance as a function of frequency for the sliding finger under the nominal (magenta) and wet (cyan) conditions as a function of frequency. f) The change in the remaining admittance as a function of frequency for the sliding finger under the nominal (magenta) and wet (cyan) conditions. The dashed purple-colored lines are the one-decade-per-decade lines fitted to the admittance curves. The change in the g) capacitance and h) resistance at the interface between the finger skin and touchscreen under the nominal and wet conditions as a function of frequency. Our proposed circuit model for the finger in contact with a voltage-induced touchscreen under the i) nominal and j) wet conditions.}
  \label{fig:Impedance}
\end{figure*}

\subsection{Electrical Impedance Measurements}
Fig. \ref{fig:Impedance}a and b present the average magnitude and phase of the electrical impedance measurements as a function of frequency, respectively. The shaded regions around the curves represent the standard error of means. Green, black, magenta, and cyan-colored curves show the electrical impedance measurements for skin, touchscreen, sliding finger in nominal condition, and sliding finger in wet condition, respectively. The individual trials of electrical impedance measurements are also presented in Fig. S10-S12 (Supplementary Material) for skin, Fig. S13-S17 (Supplementary Material) for touchscreen, Fig. S18-S20 (Supplementary Material) for sliding finger in nominal condition, and Fig. S21-S23 (Supplementary Material) for sliding finger in wet condition. 
As shown in Fig. \ref{fig:Impedance}c, we measured the apparent contact area of the participant's fingerpad using a high-resolution camera as 130 mm$^2$ for a normal force of 1 N using the approach given in our earlier study \cite{ozdamar2020step}. The apparent contact area increased to 300 mm$^2$ when the interface between the finger and the touchscreen was filled with NaCl in the wet condition (Fig. \ref{fig:Impedance}d). Note that this large value includes the area of the fingerpad plus the area of NaCl around it. Since the impedance measurements are affected by the contact area \cite{grimnes2015bioimpedance}, we normalized the electrical impedance for the wet condition by multiplying its real and imaginary parts by the ratio of 300/130.

As shown in Fig. \ref{fig:Impedance}a, the summation of the magnitudes of skin and touchscreen impedances is not equal to the magnitude of the total sliding impedance for both the nominal and wet conditions. Hence, another impedance must be in series with the skin and touchscreen impedances, which we name the "remaining impedance". We subtract the skin ($Z_{Skin}$) and touchscreen ($Z_{TS}$) impedances from the total sliding impedance ($Z_{Sliding}$) to obtain the remaining impedance \cite{shultz2018electrical, aliabbasi2024experimental}:

\begin{equation}\label{eq:subtraction}
    Z_{R} = Z_{Sliding}-Z_{Skin}-Z_{TS}
\end{equation}

The magnitude (solid curves) and phase (dashed curves) of the remaining impedance for the sliding finger under the nominal (magenta-colored curve) and wet (cyan-colored curve) conditions are presented in Fig. \ref{fig:Impedance}e as a function of frequency. The magnitude of the remaining impedance for the nominal condition was significantly higher than that of the wet condition. In other words, the liquid at the interface of the finger and the touchscreen caused a drop in impedance magnitude of more than tenfold compared to the nominal condition. As shown in Fig. \ref{fig:Impedance}e, the phase angles of the remaining impedances for the nominal and wet conditions showed a resistive behavior at lower frequencies. This resistive behavior was followed by purely capacitive behavior after approximately 30 Hz for the nominal condition (i.e. the phase angle is around -90 degrees after 30 Hz). However, the phase angle of the wet condition showed a capacitive behavior for a narrow range of frequencies, followed by a sharp return to the resistive behavior.

Fig. \ref{fig:Impedance}f shows the remaining admittances for the nominal and wet conditions. A one-decade-per-decade line (dashed purple-colored) fits well to the admittance curve of the nominal condition after 30 Hz, suggesting a constant capacitance after that frequency (Fig. \ref{fig:Impedance}g). Hence, the remaining impedance of the nominal condition can be modeled by a single capacitance ($C_{gap}$) at higher frequencies, representing the air gap between the finger and the touchscreen \cite{aliabbasi2024experimental}. At frequencies lower than 30 Hz, there is a parasitic capacitance. This capacitance occurs due to the electric double layer at the interface between the finger and the touchscreen \cite{kuang1998low}. The formation of the electric double layer at the interface of the finger surface underlies the phenomenon of electrode polarization \cite{schwan1968electrode}. Upon contact with the voltage-induced touchscreen, the human finger triggers the movement of ions in the finger tissue toward the surface of the touchscreen, causing the formation of a first layer on the inner finger surface. This primary layer consists of ions carrying electric charges opposite to those of the touchscreen, while the subsequent layer comprises loosely anchored ions bearing similar charges. The presence of free ions in the finger, possessing charges contrary to those of the touchscreen, results in their attraction towards the touchscreen, displacing the ions in the first layer and leading to the leakage of electrons from the finger to the touchscreen surface. It is worth noting that the transfer of ions from the finger to the touchscreen is restricted because the touchscreen has only electronic charge carriers rather than ionic ones. Particularly at lower frequencies, the sufficient duration allows the free ions to displace those in the first layer more effectively, leading to an increased rate of charge leakage and, subsequently, a reduction in the strength of the electric field at the interface (a conduction path builds up between the finger and the touchscreen as emerged in the remaining resistance curve in Fig. \ref{fig:Impedance}h). Hence, a capacitance ($C_{EP}$) in parallel with a resistance ($R_{EP}$) can be used to model the behavior of the electrode polarization impedance as suggested in \cite{aliabbasi2024experimental} (see Fig. \ref{fig:Impedance}i for our proposed circuit model).

Similar to the nominal condition, a one-decade-per-decade line fits well to the admittance curve of the wet condition in Fig. \ref{fig:Impedance}f for frequencies ranging from 100 Hz to 10 kHz. Hence, the remaining impedance of the wet condition can be modeled by a set of parallel capacitances for the air gap ($C^{\prime}_{gap}$) and NaCl ($C_{NaCl}$) at frequencies ranging from 100 Hz to 10 kHz. At frequencies lower than 100 Hz, there is a parasitic capacitance due to the electric double layer ($C^{\prime}_{EP}$) in parallel with a resistance ($R^{\prime}_{EP}$). As shown in Fig. \ref{fig:Impedance}h, the resistance of the wet condition is lower than that of the nominal condition. This indicates that NaCl fills the air gap and creates a conduction path between the finger and the touchscreen, which can be modeled by a resistance ($R_{NaCl}$). Fig. \ref{fig:Impedance}j shows our proposed circuit model for the wet condition. Note that the values of $C^{\prime}_{gap}$, $C^{\prime}_{EP}$, and $R^{\prime}_{EP}$ differ from the corresponding ones used for the nominal condition.  

\subsection{Overall Discussion}
We investigated the effect of input voltage signal type (DC vs. AC) on the tactile perception of EA. The earlier studies \cite{bau2010teslatouch,vardar2017effect} investigated the human tactile threshold under AC stimulation but did not correlate it with the moisture level of participants. Moreover, the results of our tactile detection experiment showed that the threshold voltage under the AC stimulation was significantly lower than that of the DC stimulation for all participants (Fig. \ref{fig:Introduction}a and b). In fact, there was no convergence in threshold voltages under the DC stimulation.

These results are consistent with our earlier findings \cite{vardar2017effect} suggesting that the Pacinian channel is mainly responsible for the tactile perception of EA at the frequency of our stimulation (125 Hz). Since the rapidly adapting receptors are not stimulated when a DC stimulation is applied to the touchscreen, it is not surprising that the magnitude of the perceived tactile stimulus is reduced. Our threshold experiment showed that the detection of EA stimuli depends not only on the amplitude of the voltage applied to the touchscreen but also on the human psychophysical sensitivity to tactile stimuli as reported in \cite{vardar2017effect}. Each psychophysical channel is sensitive to different input frequencies, which partially overlap. In our threshold experiments, the frequency of the input voltage under the AC stimulation was 125 Hz, which was primarily detected by the Pacinian channel at 250 Hz. Since the electrostatic force is proportional to the square of the input voltage, the frequency of the output force signal is twice the frequency of the input voltage applied to the touchscreen \cite{aliabbasi2022frequency}. As shown in Fig. S1c (Supplementary Material), the FFT analysis of the tangential force under AC stimulation showed a peak at 250 Hz. The FFT magnitude of the peak in the 3\textsuperscript{rd} trial (Fig. S1c, Supplementary Material) is already high due to the high voltage applied to the touchscreen, while the one in the 12\textsuperscript{th} trial (Fig. S1d, Supplementary Material) is lower in magnitude than those of some other frequencies. However, it is known that human vibrotactile perception is frequency-dependent, with a sensitivity peak around 250 Hz. If the energy contained by each frequency component is multiplied by the inverse of the normalized human sensitivity curve \cite{vardar2017effect}, then one can more easily appreciate why the small peak observed at 250 Hz in the 12\textsuperscript{th} trial affects the participant’s tactile perception significantly. This argument is also supported by the fact that the participant has successfully given a correct response under AC stimulation not only in the 12\textsuperscript{th} trial but also in the following eight trials (Fig. S1a, Supplementary Material), despite the amplitudes of voltage signals applied to the touchscreen being much lower than the one, for example, applied to the touchscreen in the 7\textsuperscript{th} trial under DC stimulation (Fig. S1b, Supplementary Material). The FFT analysis of the tangential force for this trial shows no peaks at frequencies higher than 20 Hz (Fig. S1e, Supplementary Material), suggesting that the rapidly adapting receptors were not stimulated.

The significant difference in threshold values of participants under DC and AC stimulations can also be interpreted using the circuit models shown in Fig. \ref{fig:Impedance}i and j, which were developed based on the results of electrical impedance measurements. The remaining impedances for the nominal ($Z_R$) and wet ($Z^\prime_R$) conditions can be expressed in the Laplace domain as:
\begin{equation}\label{eq:RemainingNominal}
    Z_R = \frac{R_{EP}}{1 + R_{EP} \left(C_{EP}+C_{gap}\right)s}
\end{equation}
\begin{equation}\label{eq:RemainingWet}
    Z^{\prime}_R = R_{NaCl} + \frac{R^{\prime}_{EP}}{1 + R^{\prime}_{EP} \left(C^{\prime}_{EP}+C^{\prime}_{gap}+C_{NaCl}\right)s}
\end{equation}
At low and high frequencies, Equation \ref{eq:RemainingNominal} and \ref{eq:RemainingWet} reduce to Equation \ref{eq:LimitRemainingNominal} and \ref{eq:LimitRemainingWet}, respectively.
\begin{equation}\label{eq:LimitRemainingNominal}
    \begin{split}
        & \lim_{\omega \to 0} Z_R\approx R_{EP}\\
        & \lim_{\omega \to \infty} Z_R\approx 0
    \end{split}
\end{equation}
\begin{equation}\label{eq:LimitRemainingWet}
    \begin{split}
        & \lim_{\omega \to 0} Z^{\prime}_R\approx R^{\prime}_{EP}+R_{NaCl}\\
        & \lim_{\omega \to \infty} Z^{\prime}_R\approx R_{NaCl}
    \end{split}
\end{equation}

This limit analysis shows that under the nominal condition, the capacitances $C_{EP}$ and $C_{gap}$ become effectively shunted as the stimulation frequency tends towards zero (i.e., DC stimulation), diverting all the current flow towards the resistance $R_{EP}$, consequently leading to a higher amount of charge leakage through SC in comparison to that observed under AC stimulation. In other words, the interface of the finger and touchscreen is more conductive at lower frequencies due to the leakage of electrical charges from the finger to the surface of the touchscreen. At higher frequencies, the charge leakage diminishes and the behavior of the interface becomes more capacitive. Similarly, under the wet condition, the capacitances $C^{\prime}_{EP}$, $C^{\prime}_{gap}$, and $C_{NaCl}$ short out as the stimulation frequency reaches zero and all the current flows through the resistances $R^{\prime}_{EP}$ and $R_{NaCl}$. However, the resistance of the NaCl ($R_{NaCl}$) remains effective even at higher frequencies, creating a conduction path between the finger and the touchscreen. At low frequencies, $R^{\prime}_{EP}$ + $R_{NaCl}$ $<$ $R_{EP}$ since NaCl fills in the air gap between the finger and the touchscreen and reduces the resistivity as shown in Fig. \ref{fig:Impedance}h. This overall understanding further clarifies the weaker electric field observed during DC stimulation compared to AC stimulation. 

We also investigated the effect of moisture on the tactile detection threshold voltage and found that the participants with very moist fingers (S6, S7, S9) had significantly higher threshold levels than the other participants (Fig. \ref{fig:Introduction}b). For those participants, we argue that the introduction of finger sweat into the interface reduces the strength of the electric field as demonstrated by our electrical impedance measurements. Furthermore, if the friction force acting on the subject's finger was already high due to moisture, then a relatively small increase in the same force due to EA did not contribute much to his/her tactile perception. We performed experiments with one participant having relatively dry fingers and measured the friction forces between his right hand's index finger and the touchscreen under the nominal and moist conditions with and without EA (Fig. \ref{fig:Friction}a-c). Our results showed that, even in the absence of EA, the CoF experiences a five-fold increase when transitioning from the nominal to moist condition. Earlier studies in tribology literature have also reported a variation in CoF between dry and wet conditions by a factor of 1.5 to 7 \cite{sivamani2003tribological,pasumarty2011friction,zhu2011characterization}. The increase in CoF was explained by the softening of the finger due to the absorption of water, also known as plasticization, which increases the contact area and, thus, the tangential frictional force. Others attributed this change to capillary adhesion due to meniscus formation \cite{persson2008capillary,tomlinson2011understanding}, viscous shearing of liquid bridges formed between the skin and the surface \cite{dincc1991some}, and the work of adhesion due to absorbed moisture \cite{pailler2004study,pailler2009new}.

Our friction measurements showed that the magnitude of electrostatic forces due to EA is higher for the nominal condition compared to the moist condition (Fig. \ref{fig:Friction}e). In the presence of moisture, the water particles bridge the air gap between the fingertip and the touchscreen surface. This bridging effect shortens the conduction path between the finger and the touchscreen, resulting in a diminished or lowered potential difference across the gap, causing a reduction in the electrostatic force. This observation was verified by comparing the electrical impedance measurements performed with the sliding finger under the nominal and wet conditions (Fig. \ref{fig:Impedance}a and b). In the wet condition, we intentionally added some liquid (NaCl) to the interface between the finger and the touchscreen so that the interface remains lubricated throughout the frequency sweep. As shown in Fig. \ref{fig:Impedance}a, the magnitude of the electrical impedance for the wet condition dropped by more than an order of magnitude, compared to the nominal condition. This result suggests that in the presence of sweat at the interface between the finger and the touchscreen, the electrical charges can move between the finger and the touchscreen more easily, reducing the magnitude of the strength of the electric field at the gap.

\section{Conclusion}\label{sec:conclusion}
Tactile feedback on touchscreens via EA is an exciting area of research and the number of potential applications of this technology are countless \cite{basdogan2020review}. It is also expected that surface haptics via EA in the future will extend beyond the touchscreens of electronic devices and possibly become accessible on a variety of physical surfaces. These surfaces may include not only flat, curved, and flexible structures but also those made of hard or soft materials, equipped with embedded computational capabilities. Notably, advancements in new material technologies can facilitate the integration of electronic and mechanical functionalities \cite{biswas2019emerging}.

In this study, we investigated the effect of input signal type (DC vs. AC) and moisture on our tactile perception. The detection threshold for tactile stimuli under the AC stimulation was found to be significantly lower than that of the DC stimulation for all participants. The FFT analysis of the experimental data and the proposed circuit model provided a deeper understanding of the perceptual and physical mechanisms behind the difference in tactile perception of DC vs. AC stimulation. Furthermore, we noted that moisture negatively impacts the tactile perception of EA. This result is in line with our earlier experimental study \cite{sirin2019fingerpad}, which showed that the relative increase in the coefficient of friction is smaller for higher levels of fingertip skin moisture. The earlier studies in the literature have already reported that finger moisture affects the contact dynamics under tangential loading even when there is no EA \cite{andre2010fingertip,andre2011effect}. Our measurements in this study further showed that the magnitude of electrical impedance drops significantly when sweat exists at the interface between the finger and the touchscreen (see Fig. \ref{fig:Impedance}) Sweat, a more conductive substance, fills the air gap between the finger and the touchscreen and causes a reduction in magnitude of electrostatic forces.

Our future studies will investigate the effect of controlled hydration of the interface on friction between the finger and touchscreen with and without EA. Previous research in tribology, conducted without employing EA, observed an initial rise in the coefficient of friction (CoF), which subsequently decreased as the level of moisture or liquid present at the contact interface increased. \cite{adams2007friction,andre2009continuous,pasumarty2011friction,tomlinson2011understanding}. The underlying physics driving this transformation is not fully elucidated, given the intricate nature of the transition phases between dry and fully lubricated states. This complexity is particularly pronounced in the context of contacts involving nonlinear and viscoelastic materials, such as the human finger skin, which also exhibits multi-scale surface roughness. Moreover, our understanding of the impact of EA on this lubrication transition is currently limited, with only a handful of studies touching this subject to date \cite{sirin2019fingerpad,li2020electrowetting,chatterjee2023preferential}.

\section*{Acknowledgment}
C.B. acknowledges the financial support provided by the Scientific and Technological Research Council of Turkey (TUBITAK) under contract number 123E138.

\ifCLASSOPTIONcaptionsoff
  \newpage
\fi

\renewcommand*{\bibfont}{\small}
\printbibliography
\vspace{1000px}
\begin{IEEEbiography}[{\includegraphics[width=1in,height=1.25in,clip,keepaspectratio]{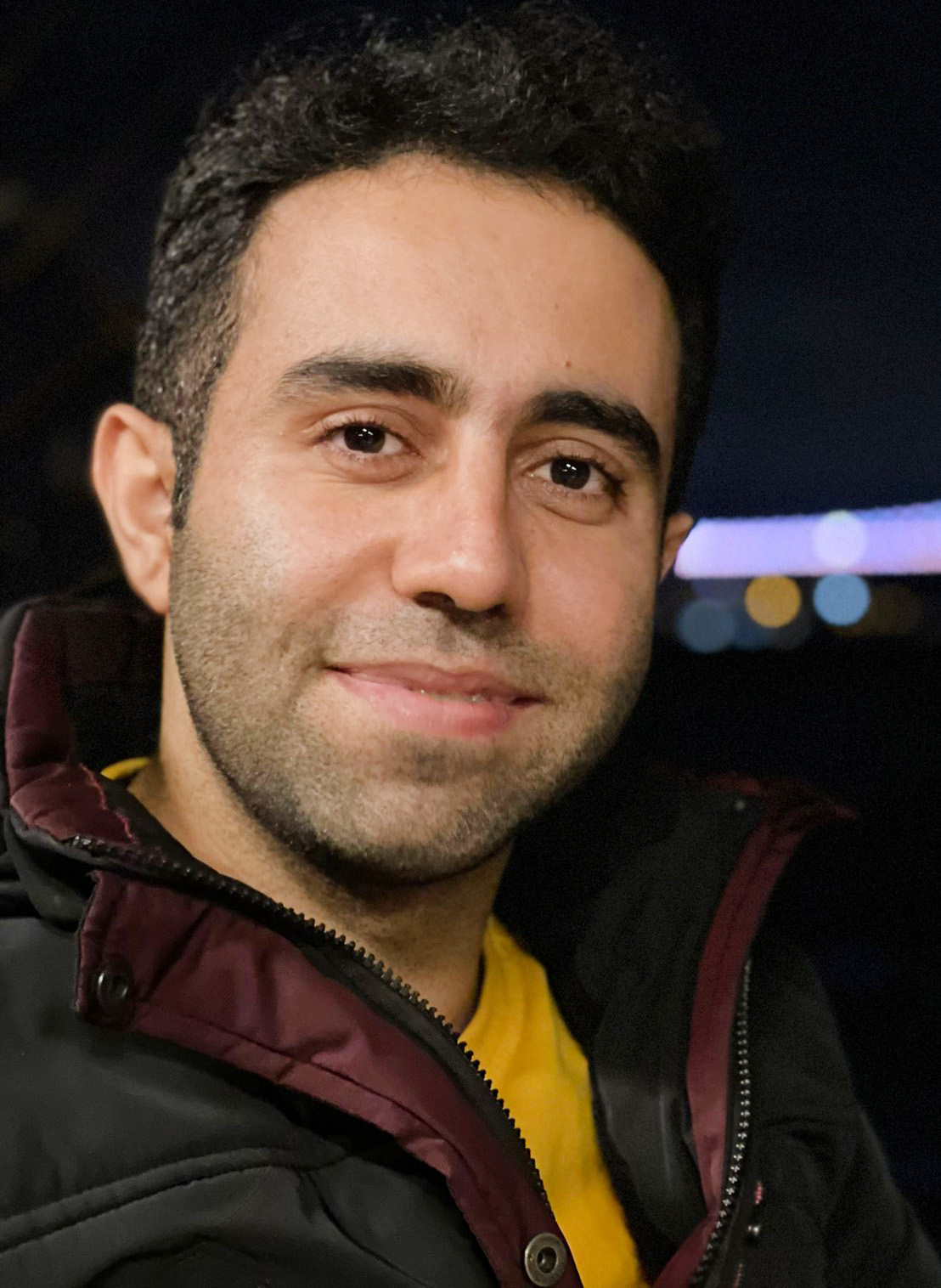}}]{Easa AliAbbasi} received the Ph.D. degree in Computational Sciences and Engineering from Koc University, Istanbul, Turkey in 2023. He is a postdoctoral researcher at Max Planck Institute for Informatics, Saarbrucken, Germany. He received his B.Sc. degree in Electrical and Electronics Engineering from Azad University in 2014 and M.Sc. degree in Mechatronics Engineering from University of Tabriz in 2017. His research interests include haptics, human-computer interaction, mechatronics, and physics-based modeling.
\end{IEEEbiography}

\begin{IEEEbiography}[{\includegraphics[width=1in,height=1.25in,clip,keepaspectratio]{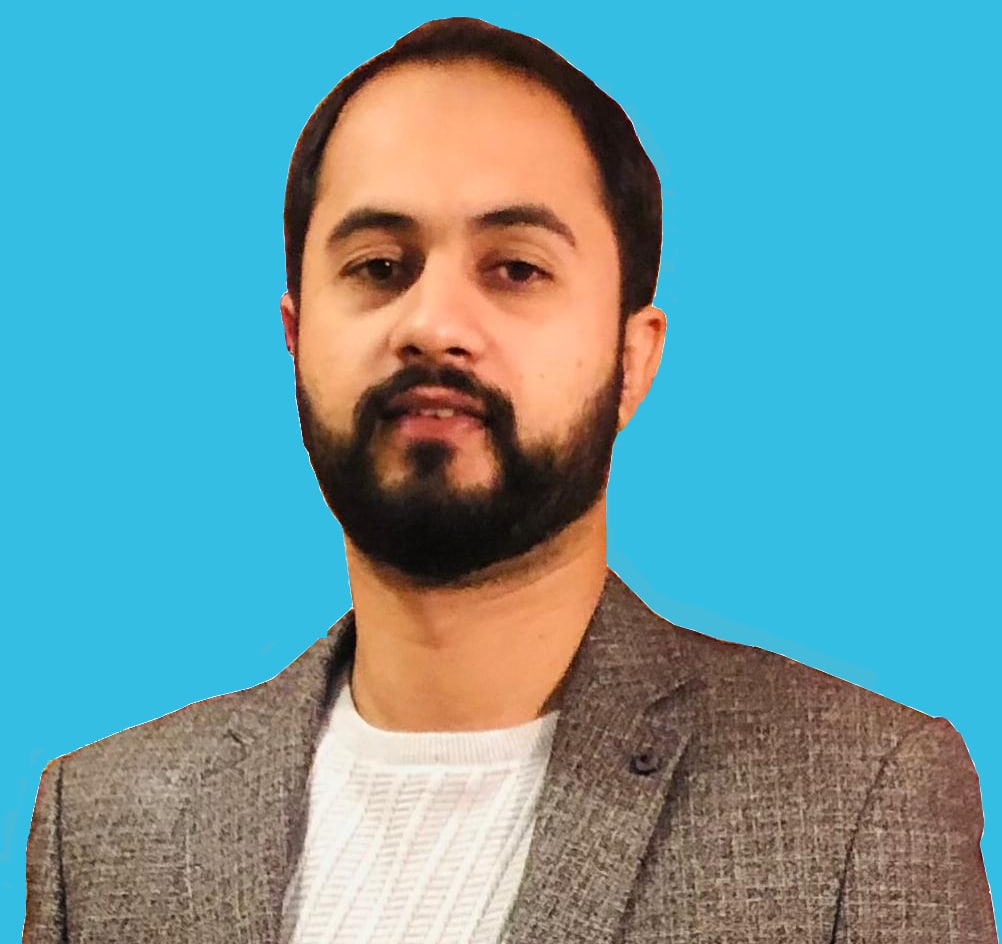}}]{Muhammad Muzammil} received the M.Sc. degree in Energetic Materials Engineering and the B.Sc. degree in Avionics Engineering from the National University of Sciences and Technology, Islamabad, Pakistan in 2019 and 2010, respectively. He is currently working toward the Ph.D. degree in Computational Sciences and Engineering program of Koc University, Istanbul, Turkey. His research interests include haptic interfaces, tactile perception, and MEMS.
\end{IEEEbiography}

\begin{IEEEbiography}[{\includegraphics[width=1in,height=1.25in,clip,keepaspectratio]{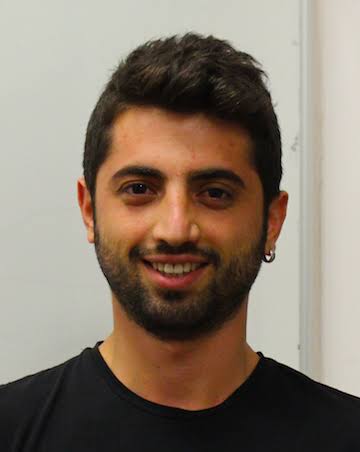}}]{Omer Sirin}
received his Ph.D. degree from Mechanical Engineering Department of Koc University in Istanbul, Turkey in 2019. He was awarded the prestigious TUBITAK BIDEB fellowship for his doctoral studies. He received his B.Sc. degree in Mechatronics Engineering from Bahcesehir University in 2012 and his M.Sc. degree in Biomedical Engineering from Cleveland State University in 2013. His research interests are haptics, mechatronics, and contact mechanics.
\end{IEEEbiography}

\begin{IEEEbiography}[{\includegraphics[width=1.25in,height=1.5in, clip,keepaspectratio, angle=-90]{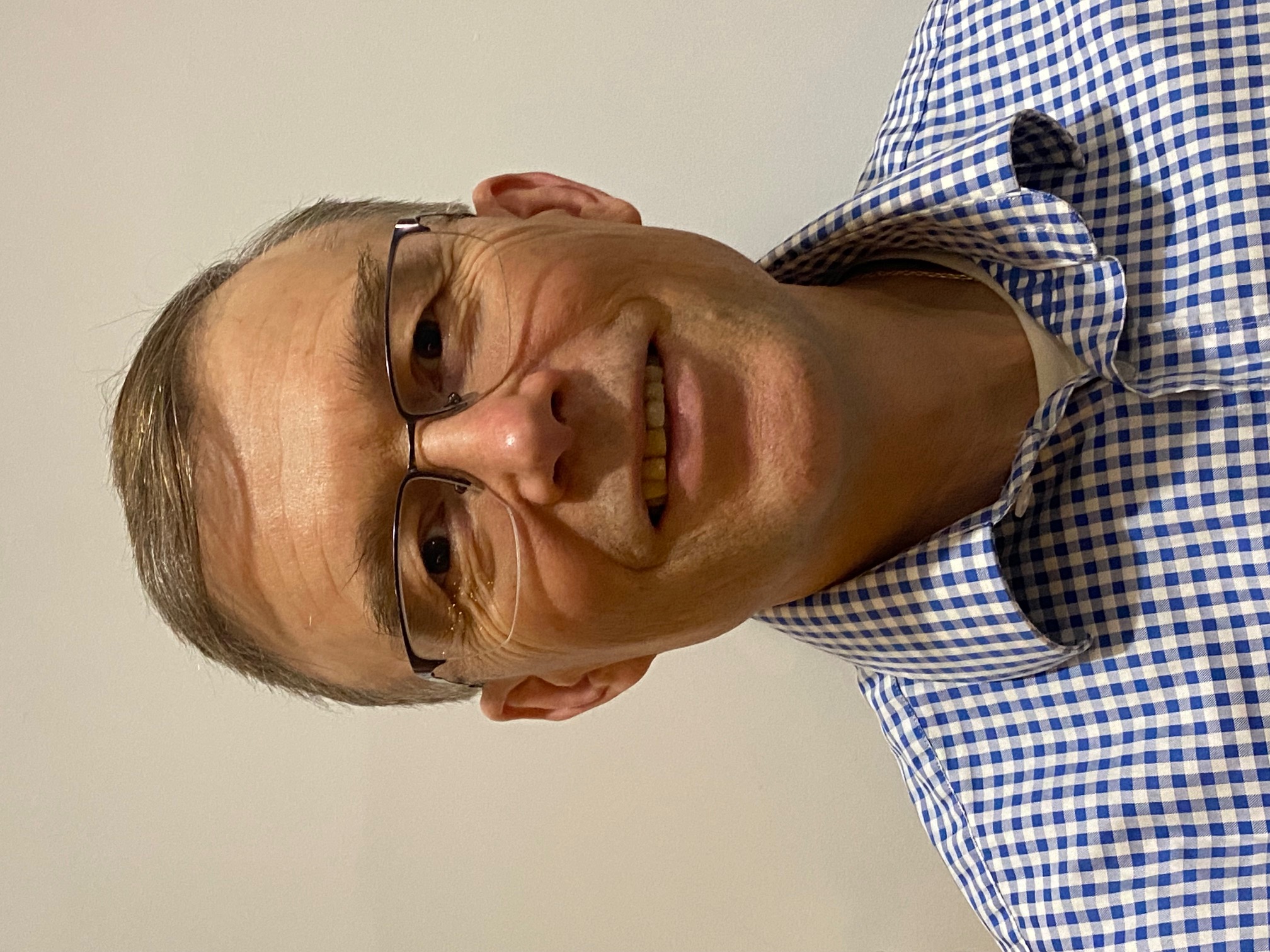}}]{Philippe Lefèvre}
graduated as an Electrical Engineer in 1988 and obtained his PhD in Applied Sciences in 1992 from UCLouvain. During his PhD, he spent one year at McGill University in the Department of Biomedical Engineering. He then spent two years (postdoc) as a Visiting Fellow at the Laboratory of Sensorimotor Research, NEI, National Institutes of Health, MD Bethesda. In 1997 he obtained a permanent position as a Research Associate from FNRS at UCLouvain. From 2003 to 2004 he spent a sabbatical and was appointed as a Visiting Scientist at the National Eye Institute, NIH, Bethesda. In 2011 he became Full Professor of Biomedical Engineering at UCLouvain. His research interests include the interaction between vision and the neural control of movement, modeling of the oculomotor and motor systems, experimental and clinical study of eye, head and limb movements, eye-hand coordination, biomechanics of finger object interaction and dexterous manipulation in micro-gravity.
\end{IEEEbiography}

\begin{IEEEbiography}[{\includegraphics[width=1in,height=1.25in, clip,keepaspectratio]{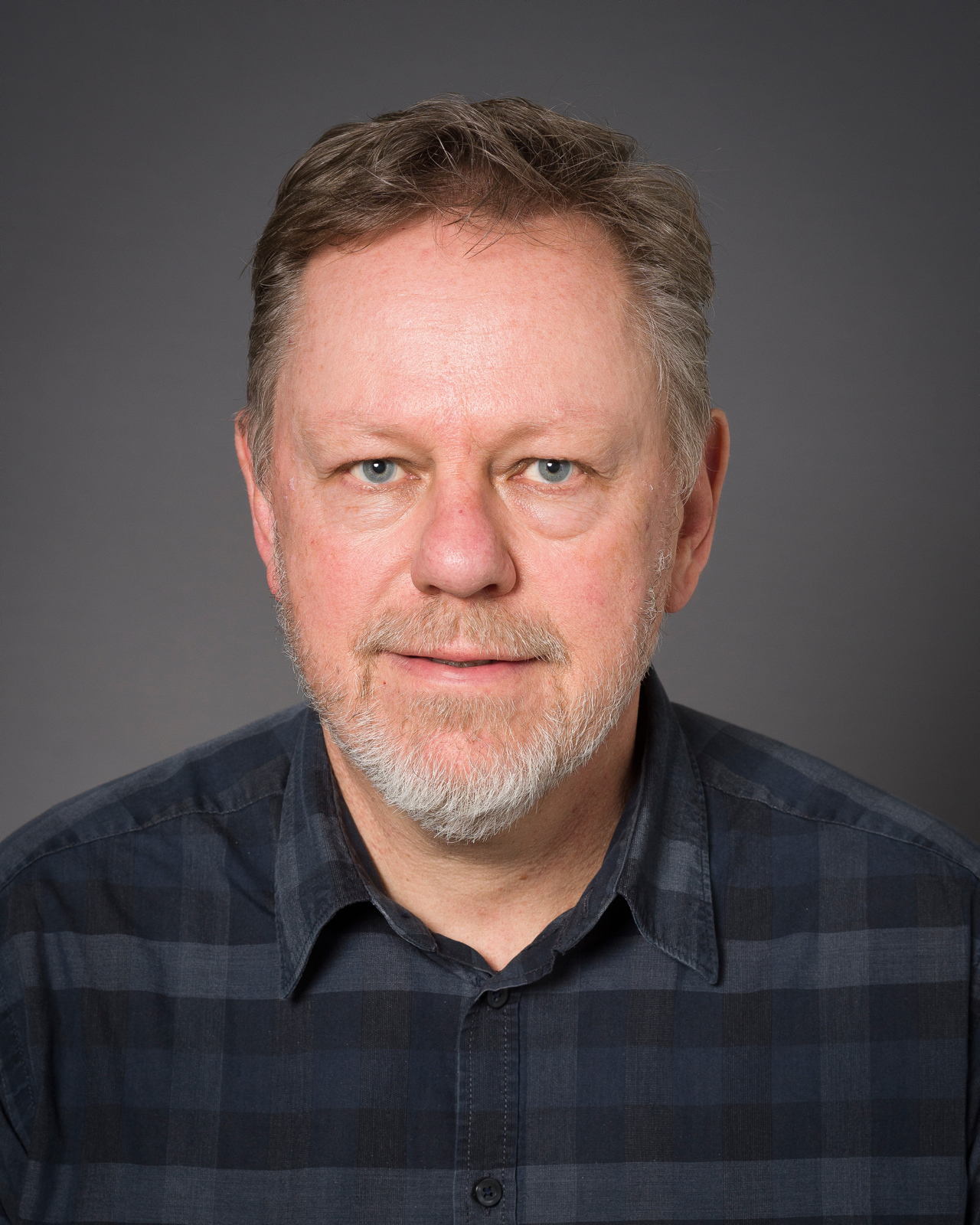}}]{Ørjan Grøttem Martinsen}
is a professor of physics and electronics at the Department of Physics, University of Oslo, Norway. He is also a senior researcher at the Oslo University Hospital. Martinsen is head of the “Oslo Bioimpedance and Medical Technology Group”, where the main focus is on electrical bioimpedance theory and applications within medicine. He is the co-author of the textbook “Bioimpedance and Bioelectricity Basics”, and he is the founding editor-in-chief of the Journal of Electrical Bioimpedance. 
\end{IEEEbiography}

\begin{IEEEbiography}[{\includegraphics[width=1in,height=1.25in, clip,keepaspectratio]{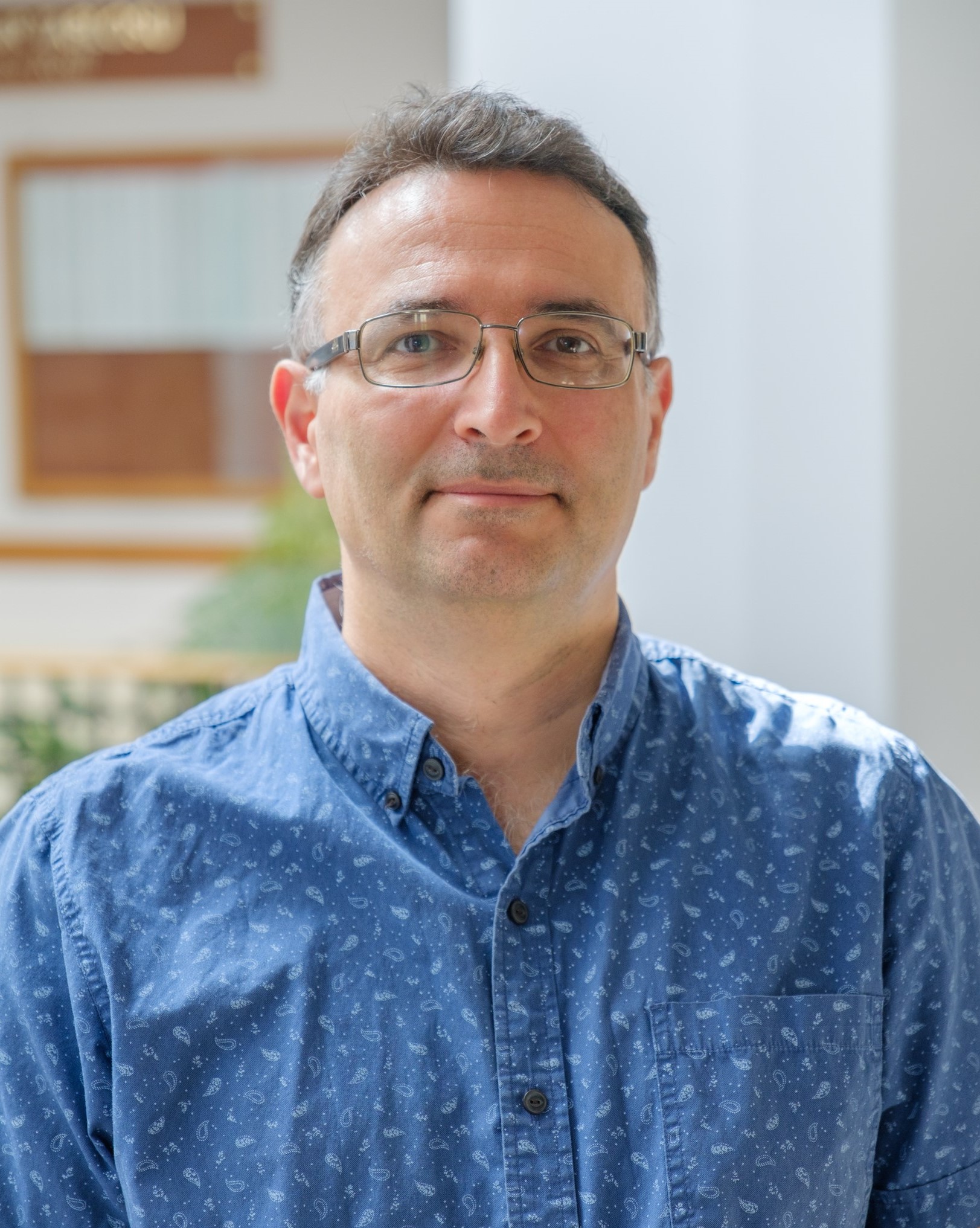}}]{Cagatay Basdogan} received the Ph.D. degree in mechanical engineering from Southern Methodist University. He is a faculty member in the mechanical engineering and computational sciences and engineering programs at the College of Engineering, Koc University, Istanbul. Before joining to Koc University, he worked at NASA-JPL/Caltech, the Massachusetts Institute of Technology, and Northwestern University Research Park. His research interests include haptic interfaces, robotics, mechatronics, biomechanics, medical simulation, computer graphics, and multi-modal virtual environments. In addition to serving in editorial boards of several journals and programme committees of conferences, he chaired the IEEE World Haptics Conference in 2011.
\end{IEEEbiography}



\section*{Supplementary material}

## SI. 1  Results of Tactile Perception Experiment

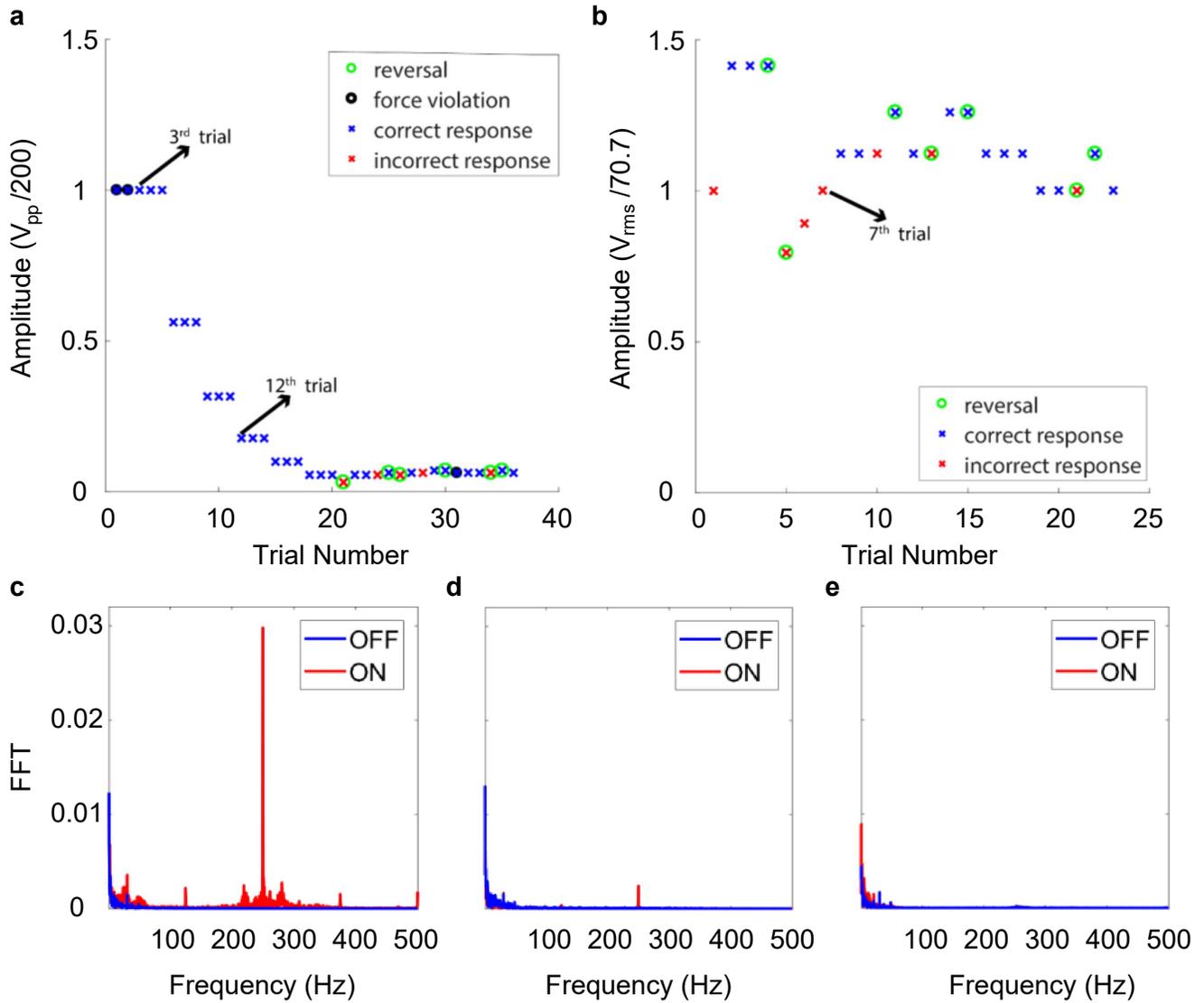

**Supplementary Figure S1.** An exemplar staircase data obtained by the threshold experiments under a) AC and b) DC conditions for participant S2. FFT analysis of the tangential force under AC condition for the c) 3rd and d) 12th trials and under DC condition for e) 7th trial.

## SI. 2  Results of Friction Measurements

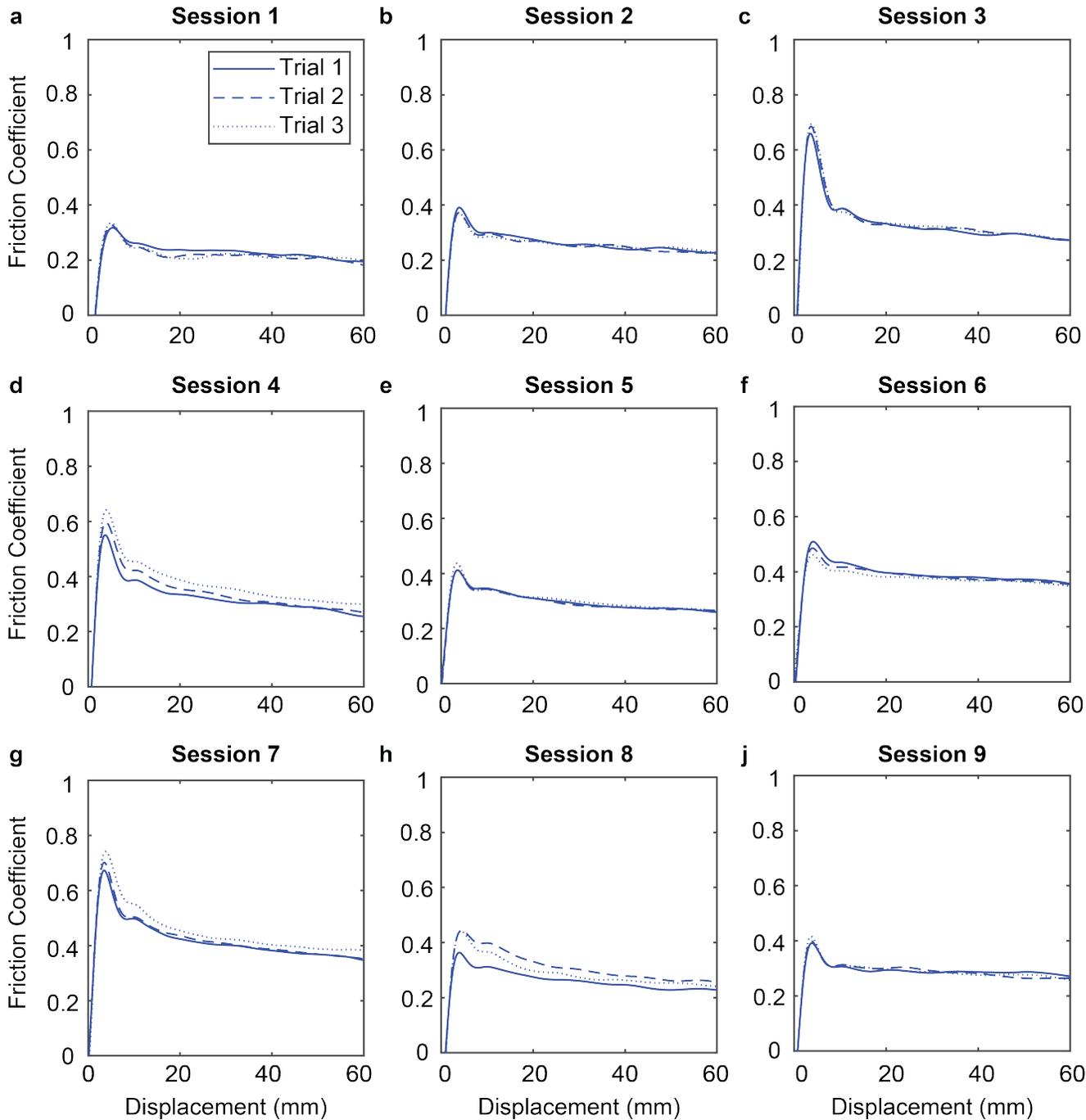

**Supplementary Figure S2.** Change in coefficient of friction (CoF) as a function of displacement for all 9 sessions (3 sessions/day x 3 days) under the nominal finger condition when EA=OFF. The solid, dashed, and dotted curves represent 1st, 2nd and 3rd trials respectively in each session.

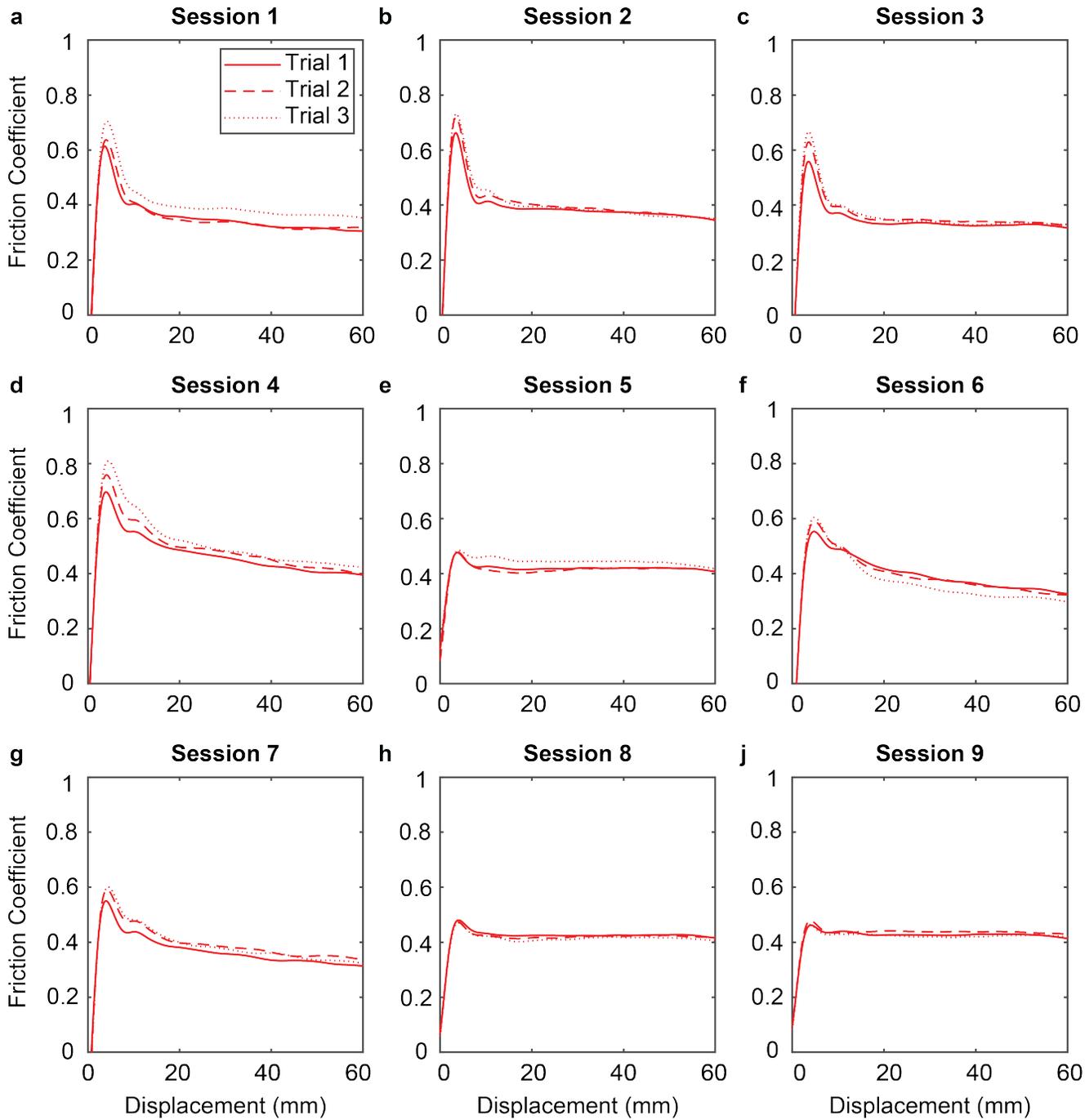

**Supplementary Figure S3.** Change in coefficient of friction (CoF) as a function of displacement for all 9 sessions (3 sessions/day x 3 days) under the nominal finger condition when EA=ON. The solid, dashed, and dotted curves represent 1st, 2nd and 3rd trials respectively in each session.

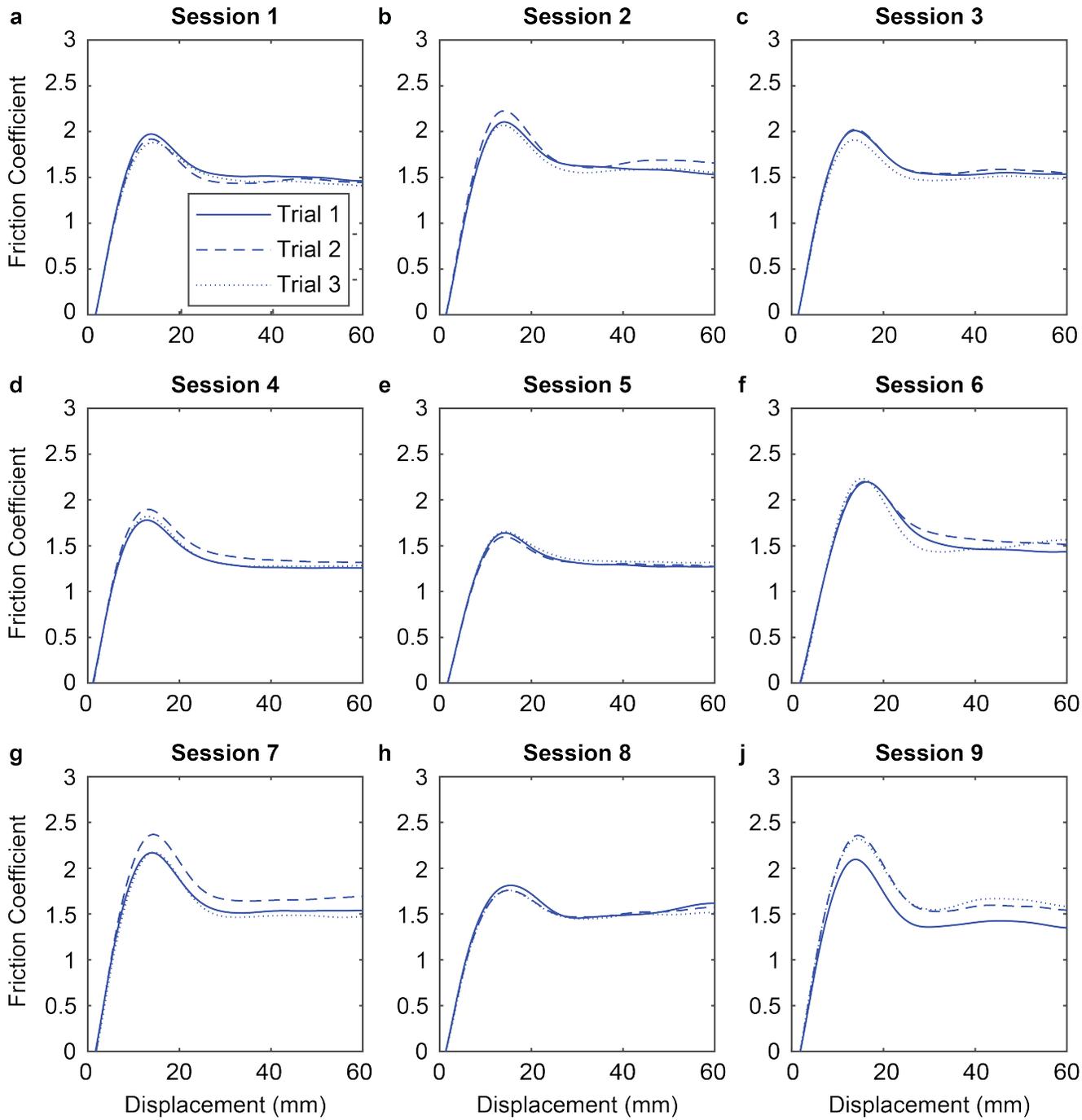

**Supplementary Figure S4.** Change in coefficient of friction (CoF) as a function of displacement for all 9 sessions (3 sessions/day x 3 days) under the moist finger condition when EA=OFF. The solid, dashed, and dotted curves represent $1^{st}$, $2^{nd}$ and $3^{rd}$ trials respectively in each session.

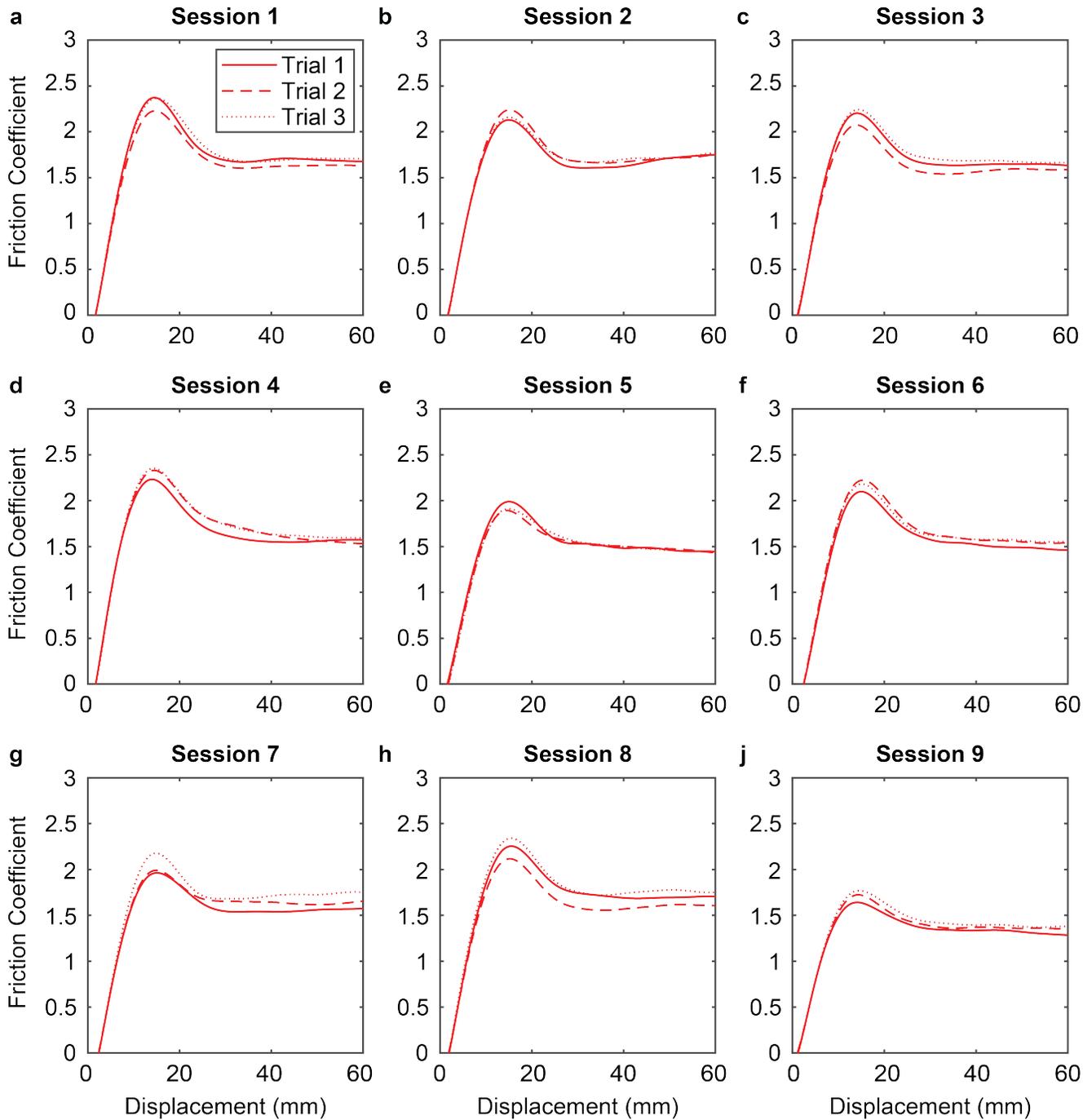

**Supplementary Figure S5.** Change in coefficient of friction (CoF) as a function of displacement for all 9 sessions (3 sessions/day x 3 days) under the moist finger condition when EA=ON. The solid, dashed, and dotted curves represent $1^{st}$, $2^{nd}$ and $3^{rd}$ trials respectively in each session.

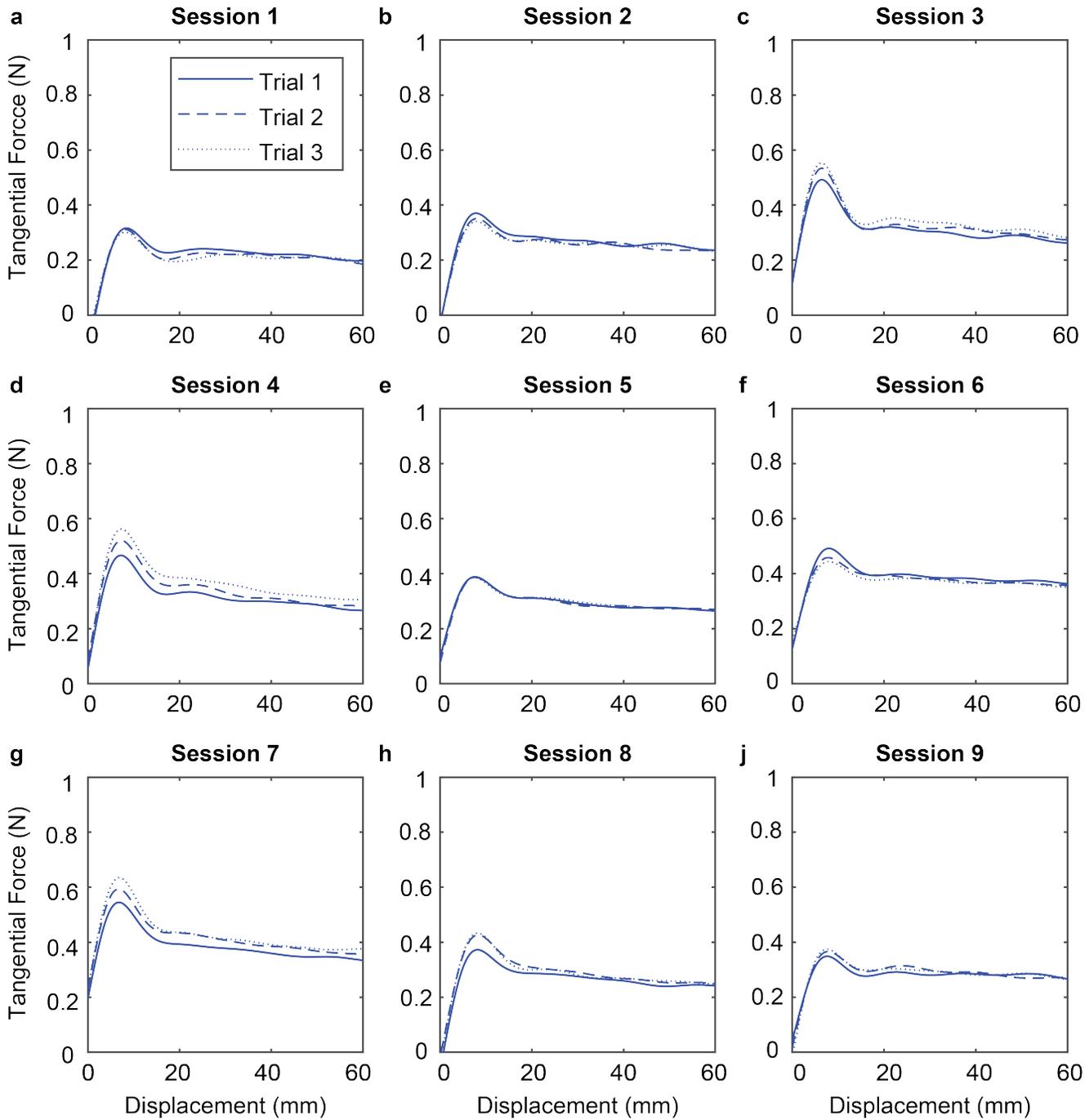

**Supplementary Figure S6.** Change in tangential force as a function of displacement for all 9 sessions (3 sessions/day x 3 days) under the nominal finger condition when EA=OFF. The solid, dashed, and dotted curves represent 1st, 2nd and 3rd trials respectively in each session.

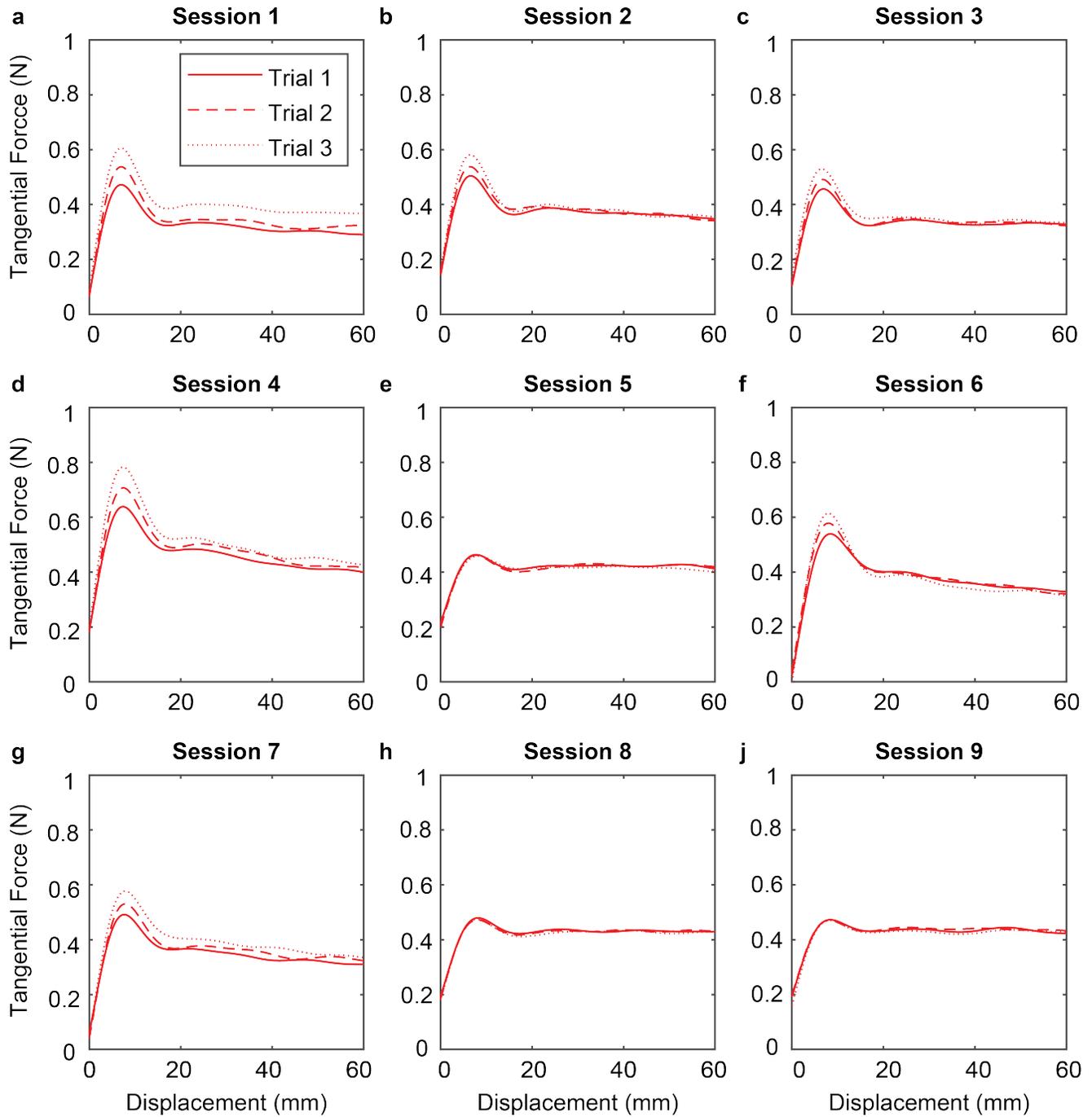

**Supplementary Figure S7.** Change in tangential force as a function of displacement for all 9 sessions (3 sessions/day x 3 days) under the nominal finger condition when EA=ON. The solid, dashed, and dotted curves represent 1st, 2nd and 3rd trials respectively in each session.

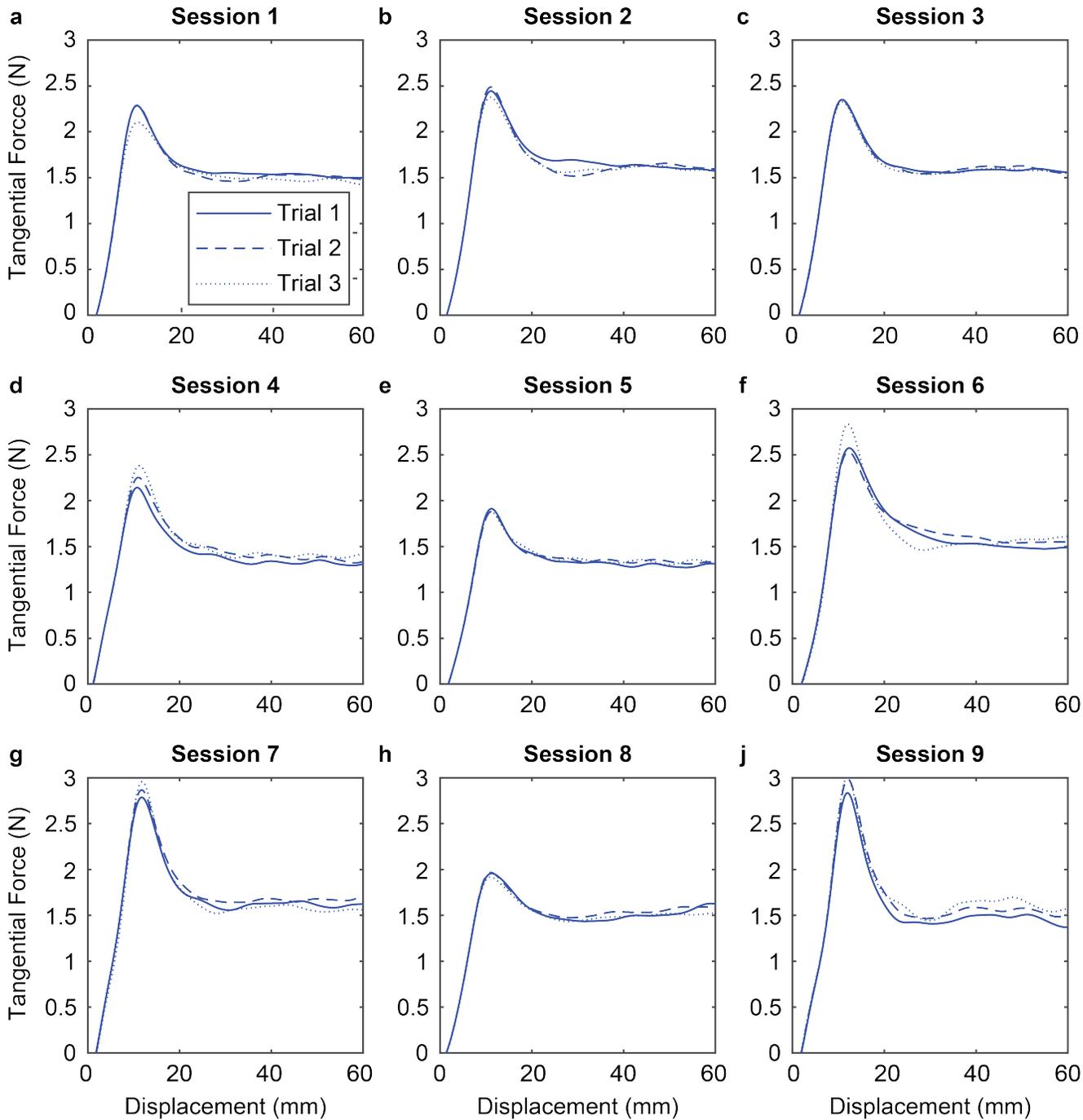

**Supplementary Figure S8.** Change in tangential force as a function of displacement for all 9 sessions (3 sessions/day x 3 days) under the moist finger condition when EA=OFF. The solid, dashed, and dotted curves represent 1st, 2nd and 3rd trials respectively in each session.

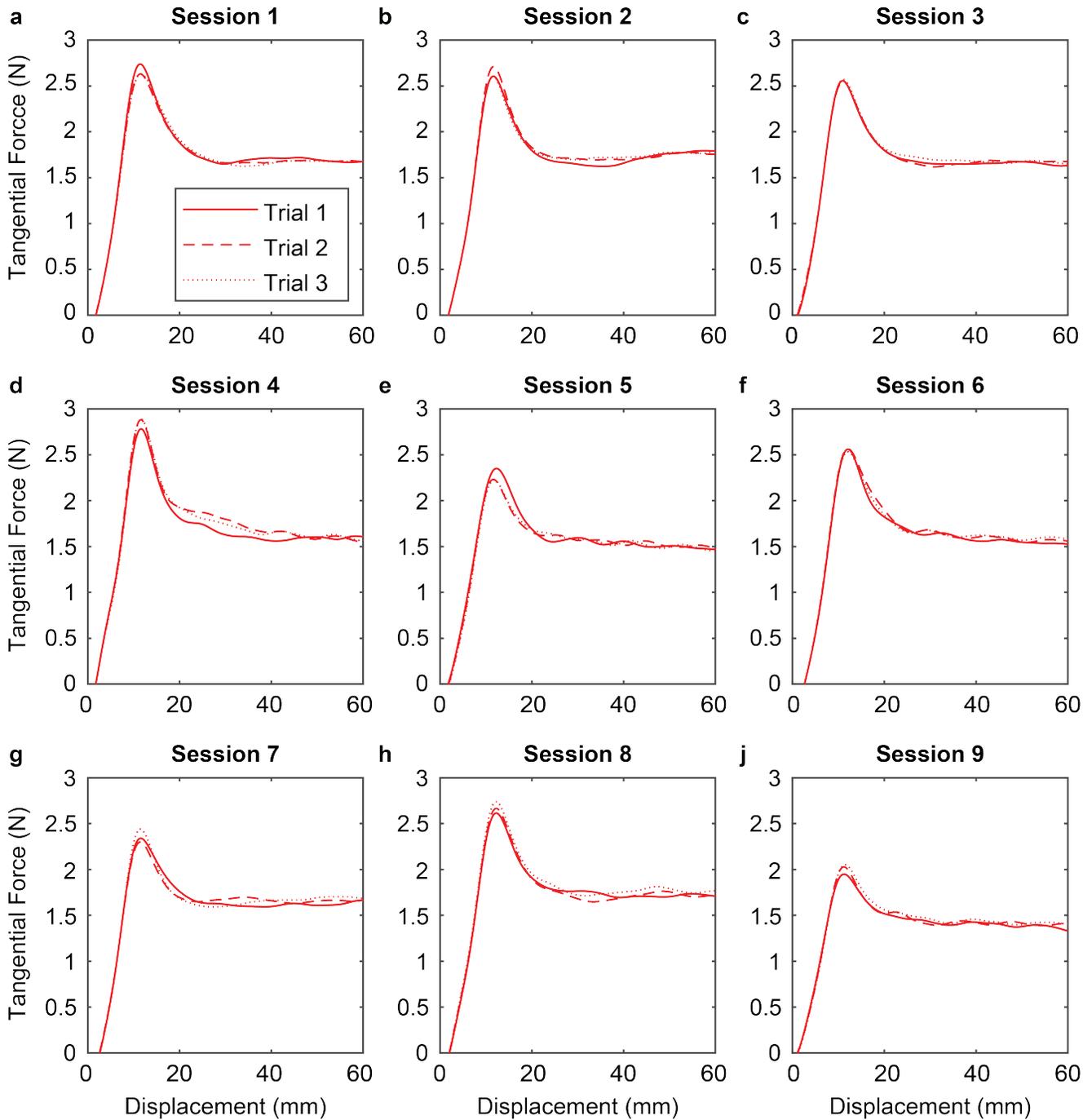

**Supplementary Figure S9.** Change in tangential force as a function of displacement for all 9 sessions (3 sessions/day x 3 days) under the moist finger condition when EA=ON. The solid, dashed, and dotted curves represent 1ˢᵗ, 2ⁿᵈ and 3ʳᵈ trials respectively in each session.

## SI. 3  Electrical Impedance Measurements for Finger Skin

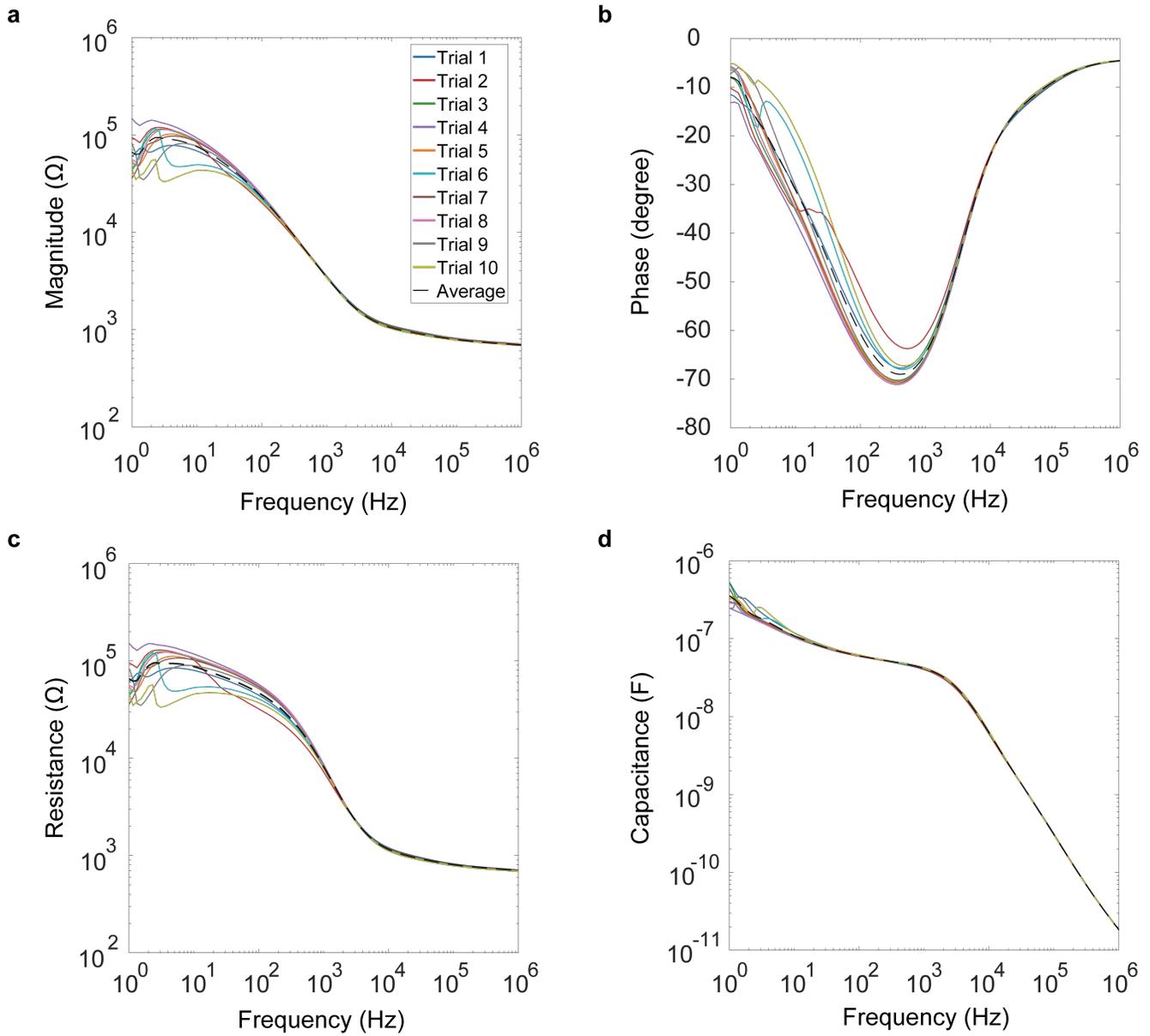

**Supplementary Figure S10.** Change in skin impedance as a function of frequency for day 1 (repeated ten times): a) magnitude, b) phase, c) resistance, and d) capacitance.

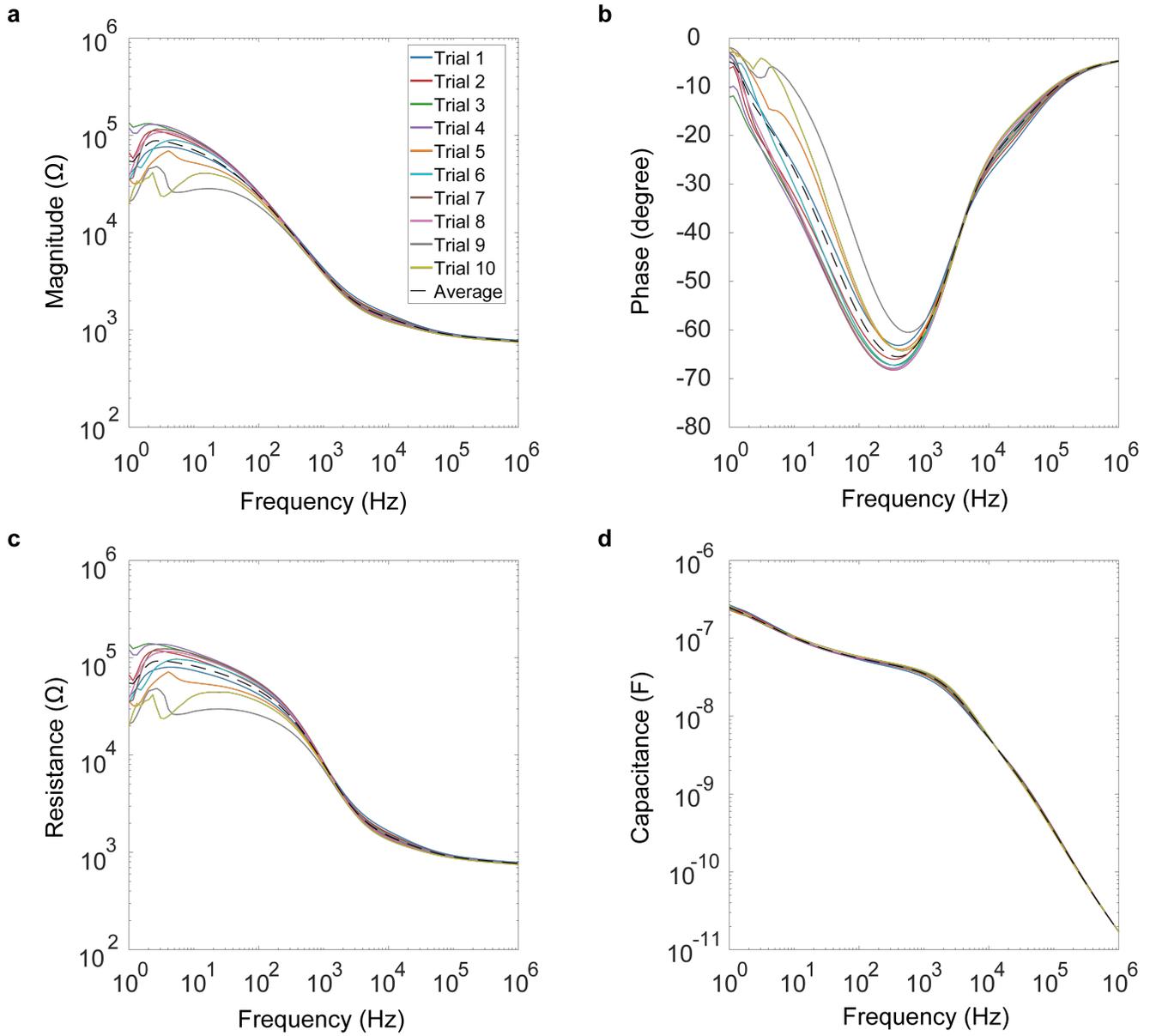

**Supplementary Figure S11.** Change in skin impedance as a function of frequency for day 2 (repeated ten times): a) magnitude, b) phase, c) resistance, and d) capacitance.

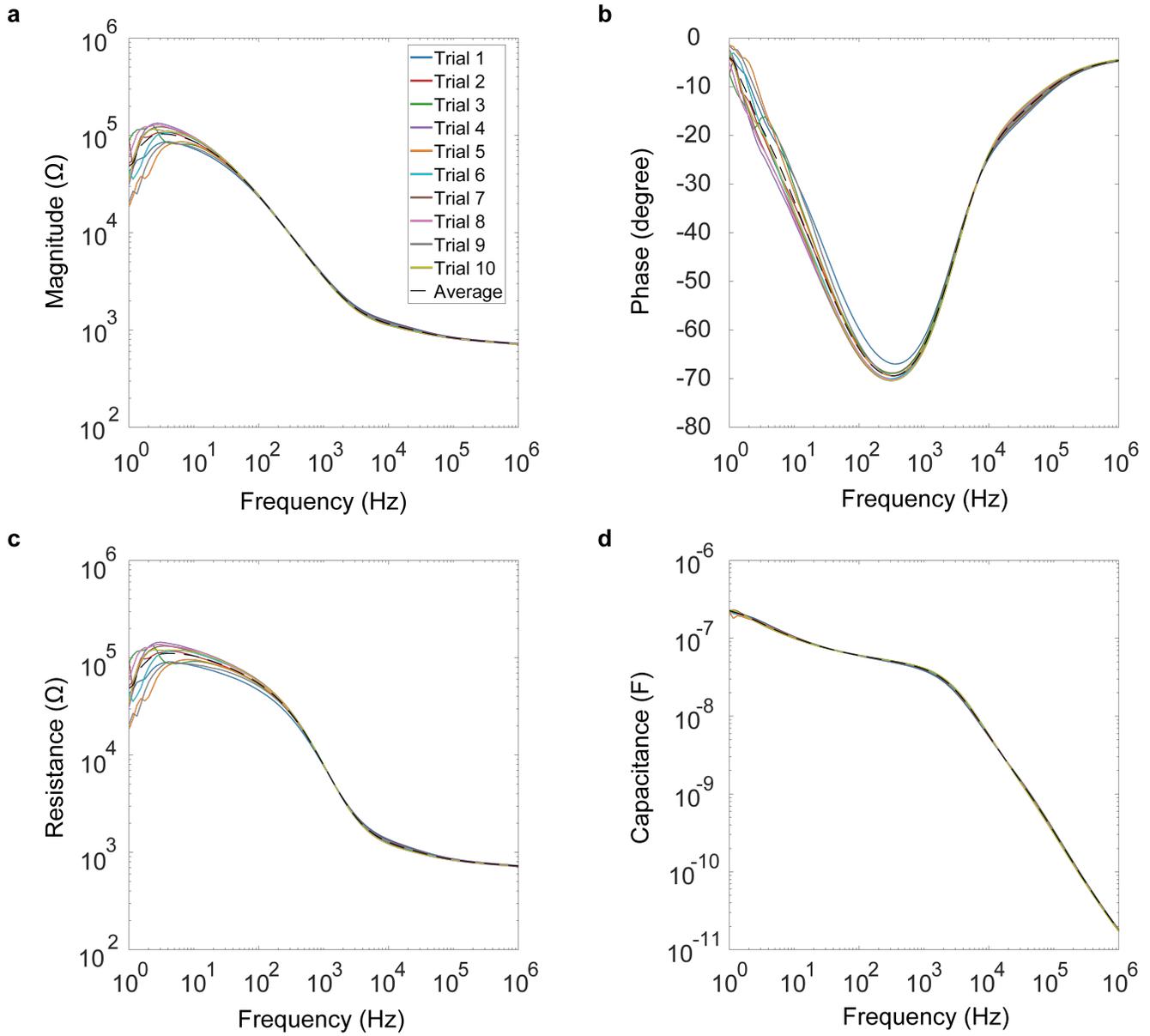

**Supplementary Figure S12.** Change in skin impedance as a function of frequency for day 3 (repeated ten times): a) magnitude, b) phase, c) resistance, and d) capacitance.

## SI. 4  Electrical Impedance Measurements for Touchscreen

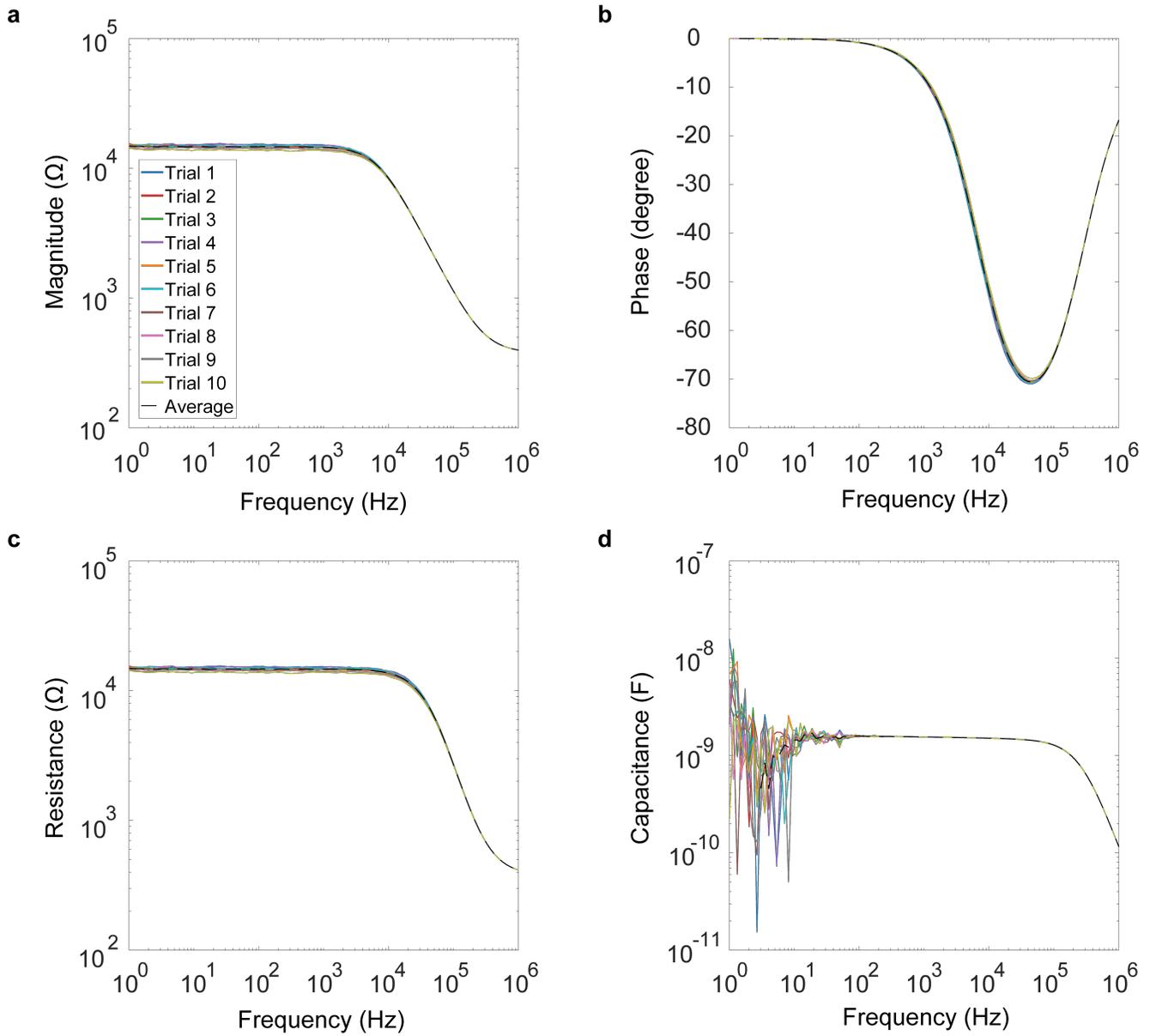

**Supplementary Figure S13.** Change in touchscreen impedance as a function of frequency, measured at location 1 (repeated ten times): a) magnitude, b) phase, c) resistance, and d) capacitance.

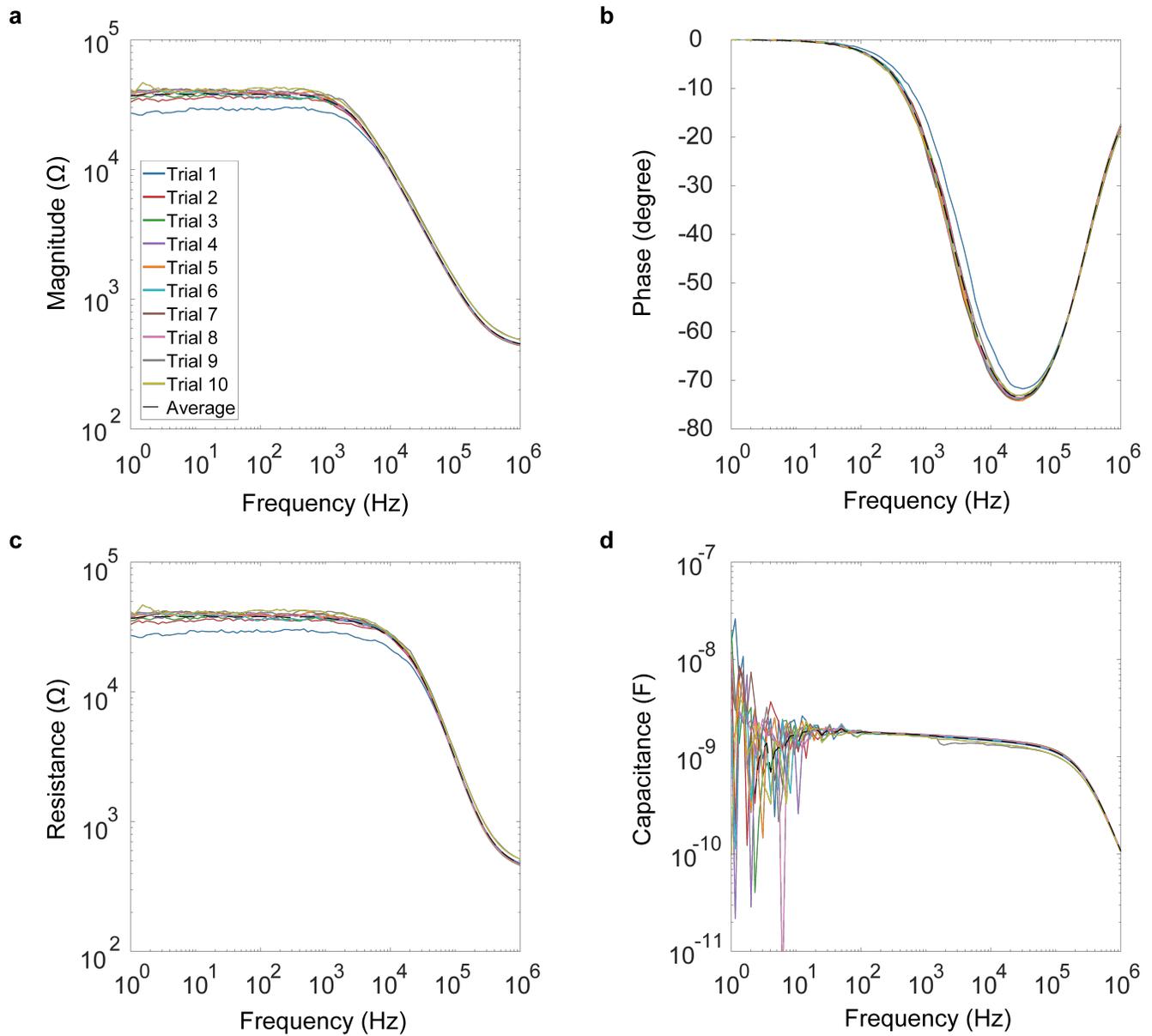

**Supplementary Figure S14.** Change in touchscreen impedance as a function of frequency, measured at location 2 (repeated ten times): a) magnitude, b) phase, c) resistance, and d) capacitance.

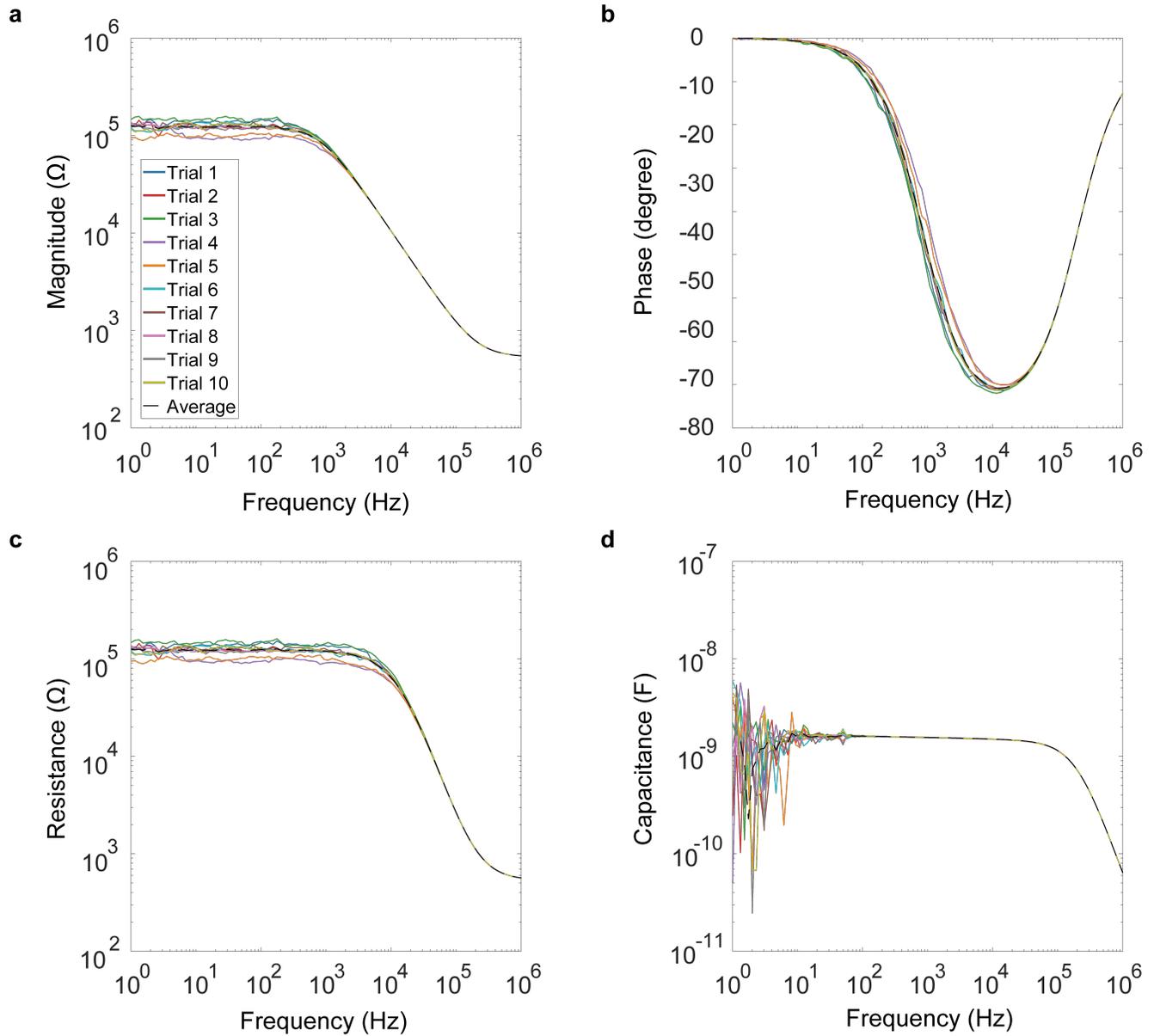

**Supplementary Figure S15.** Change in touchscreen impedance as a function of frequency, measured at location 3 (repeated ten times): a) magnitude, b) phase, c) resistance, and d) capacitance.

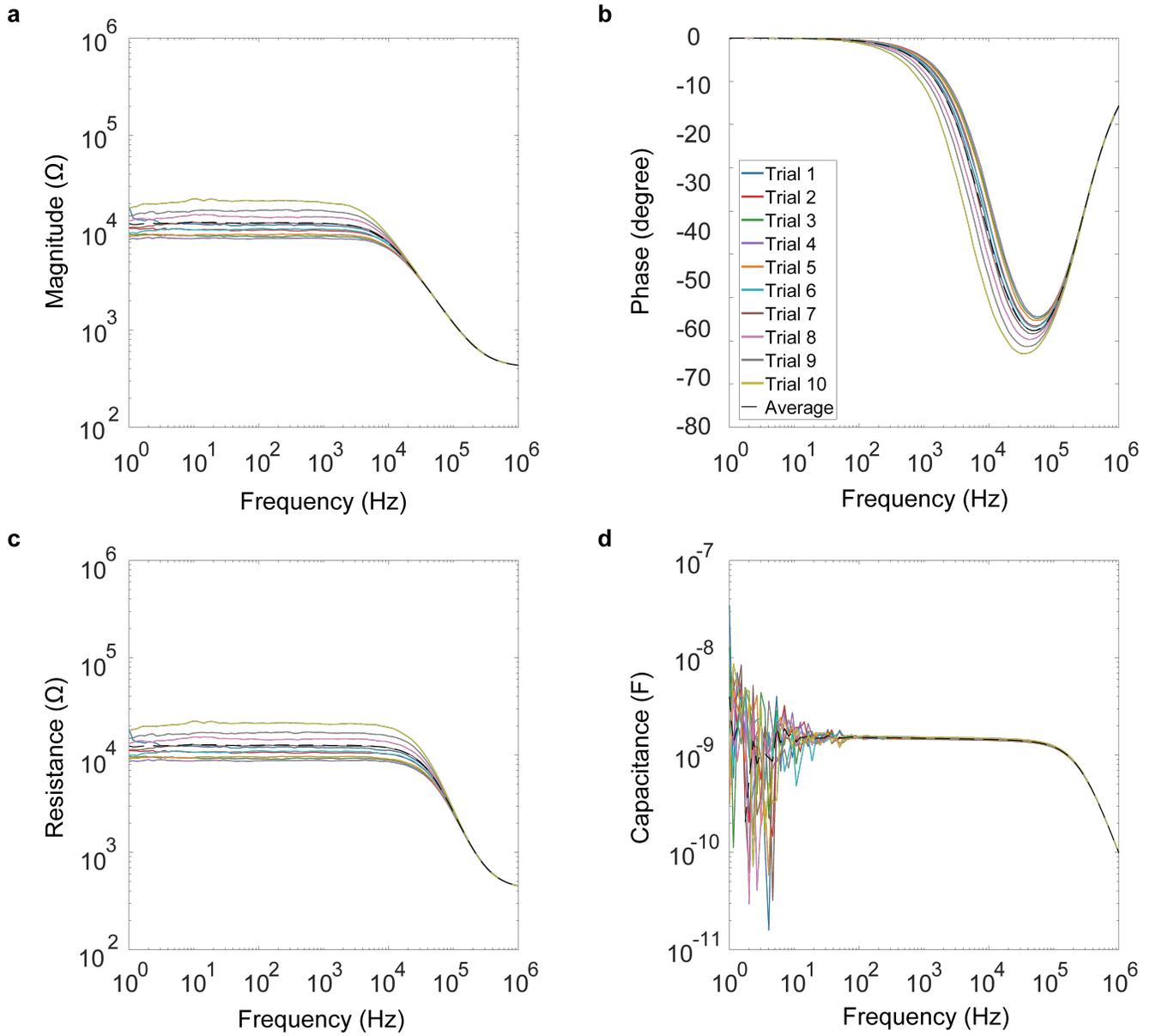

**Supplementary Figure S16.** Change in touchscreen impedance as a function of frequency, measured at location 4 (repeated ten times): a) magnitude, b) phase, c) resistance, and d) capacitance.

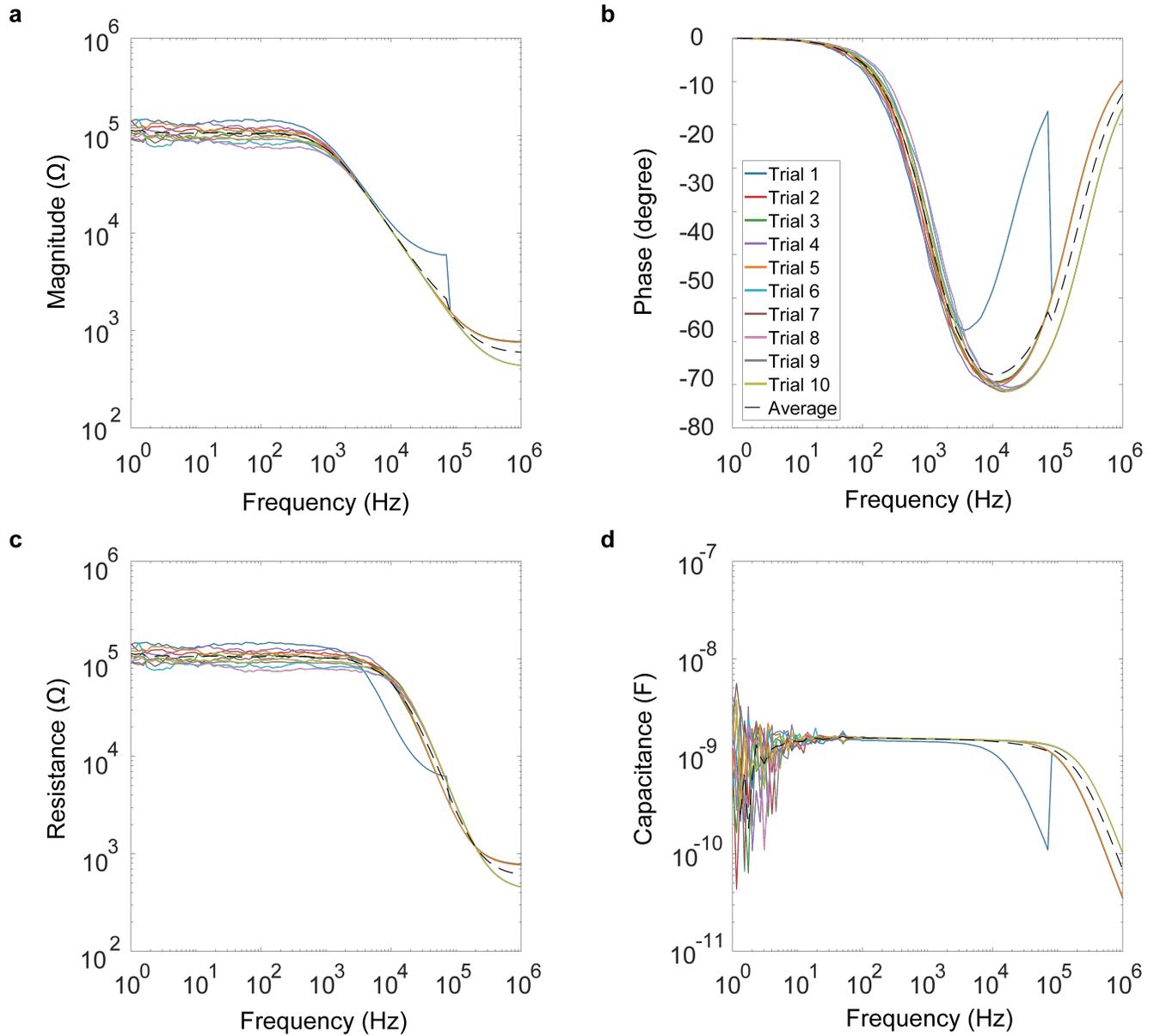

**Supplementary Figure S17.** Change in touchscreen impedance as a function of frequency, measured at location 5 (repeated ten times): a) magnitude, b) phase, c) resistance, and d) capacitance.

## SI. 5  Electrical Impedance Measurements for Finger Sliding on Touchscreen in Nominal Condition

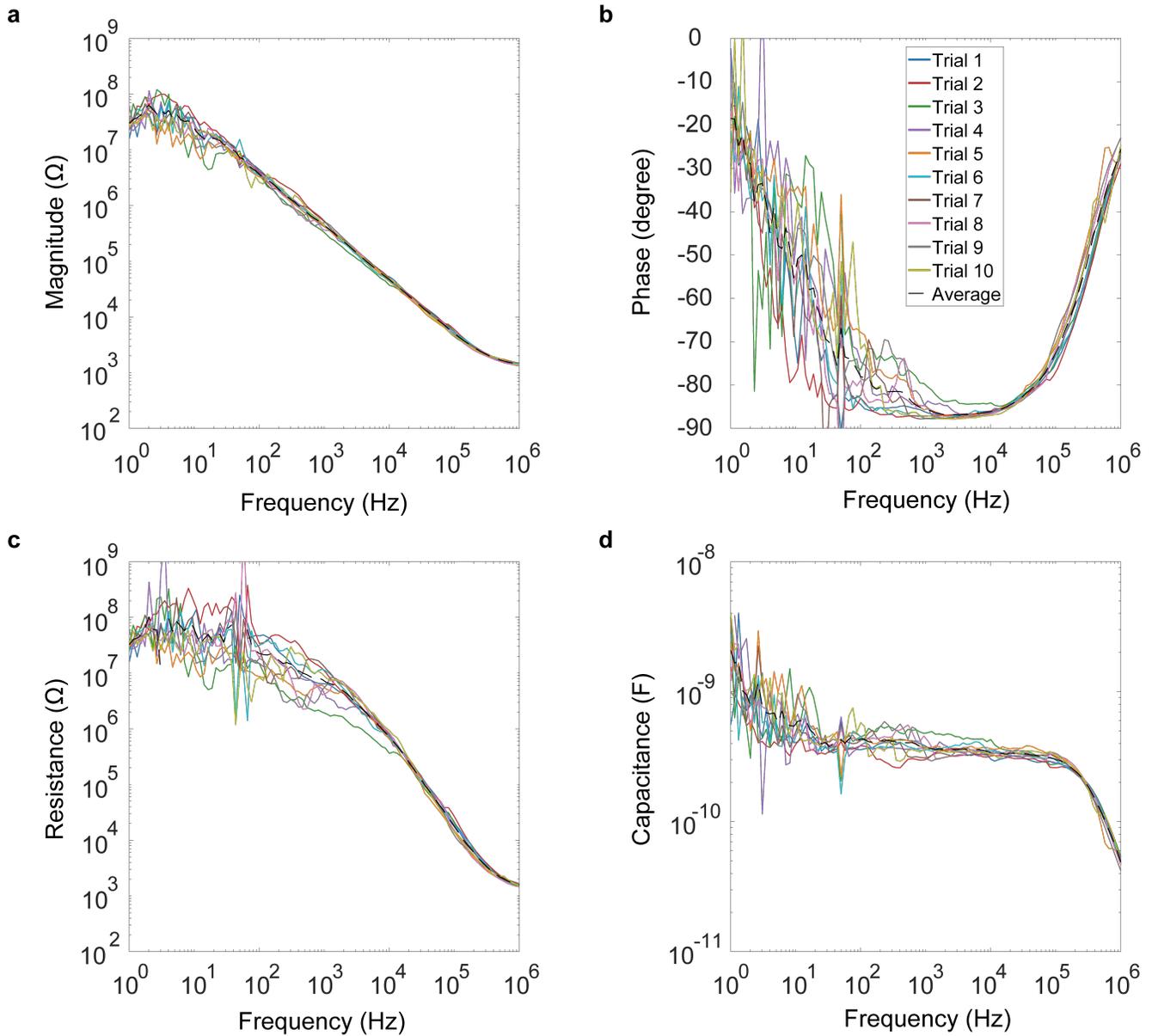

**Supplementary Figure S18.** Change in total impedance as a function of frequency for the finger sliding on the touchscreen under the nominal condition for day 1 (repeated ten times): a) magnitude, b) phase, c) resistance, and d) capacitance.

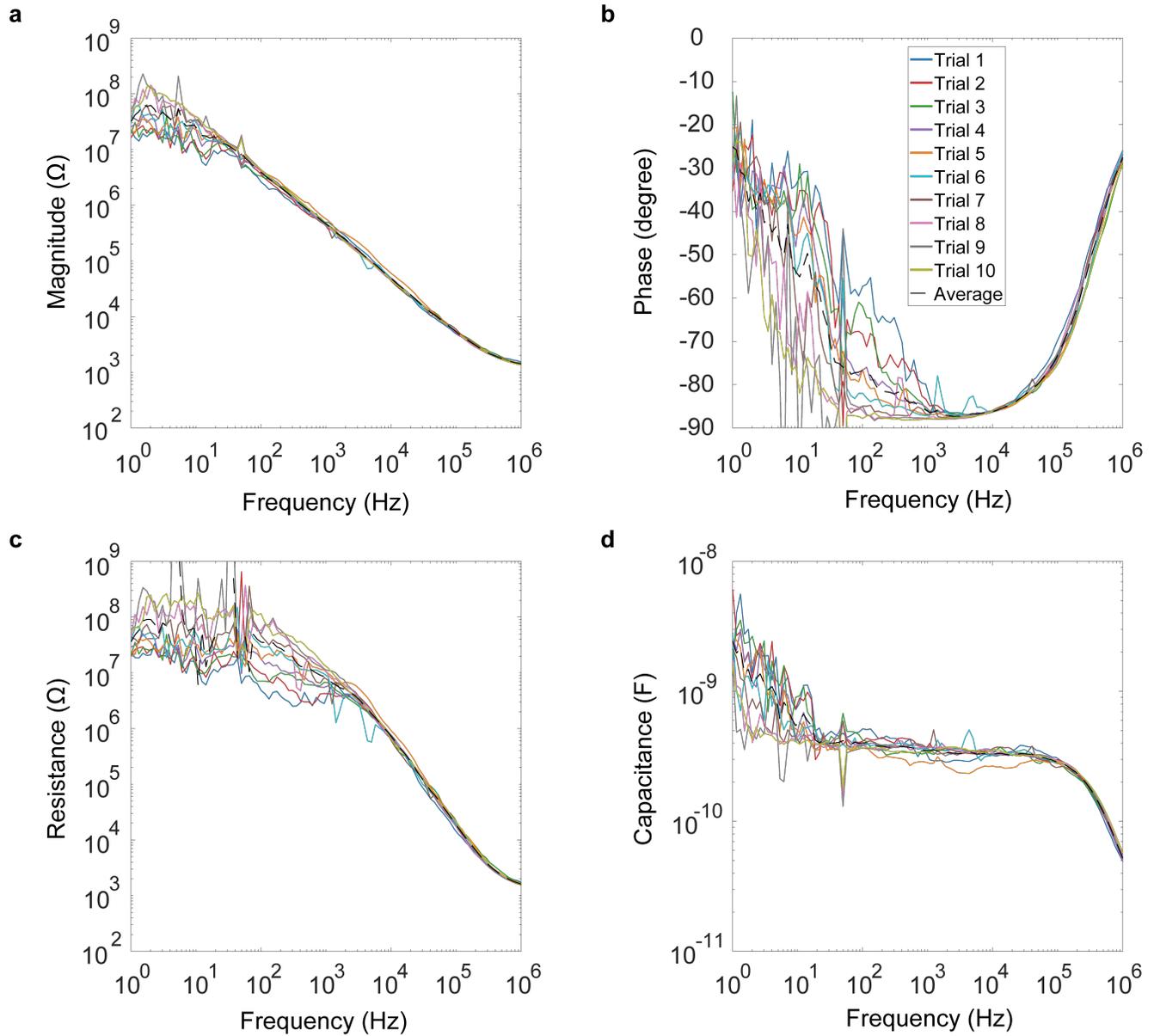

**Supplementary Figure S19.** Change in total impedance as a function of frequency for the finger sliding on the touchscreen under the nominal condition for day 2 (repeated ten times): a) impedance magnitude, b) phase, c) resistance, and d) capacitance.

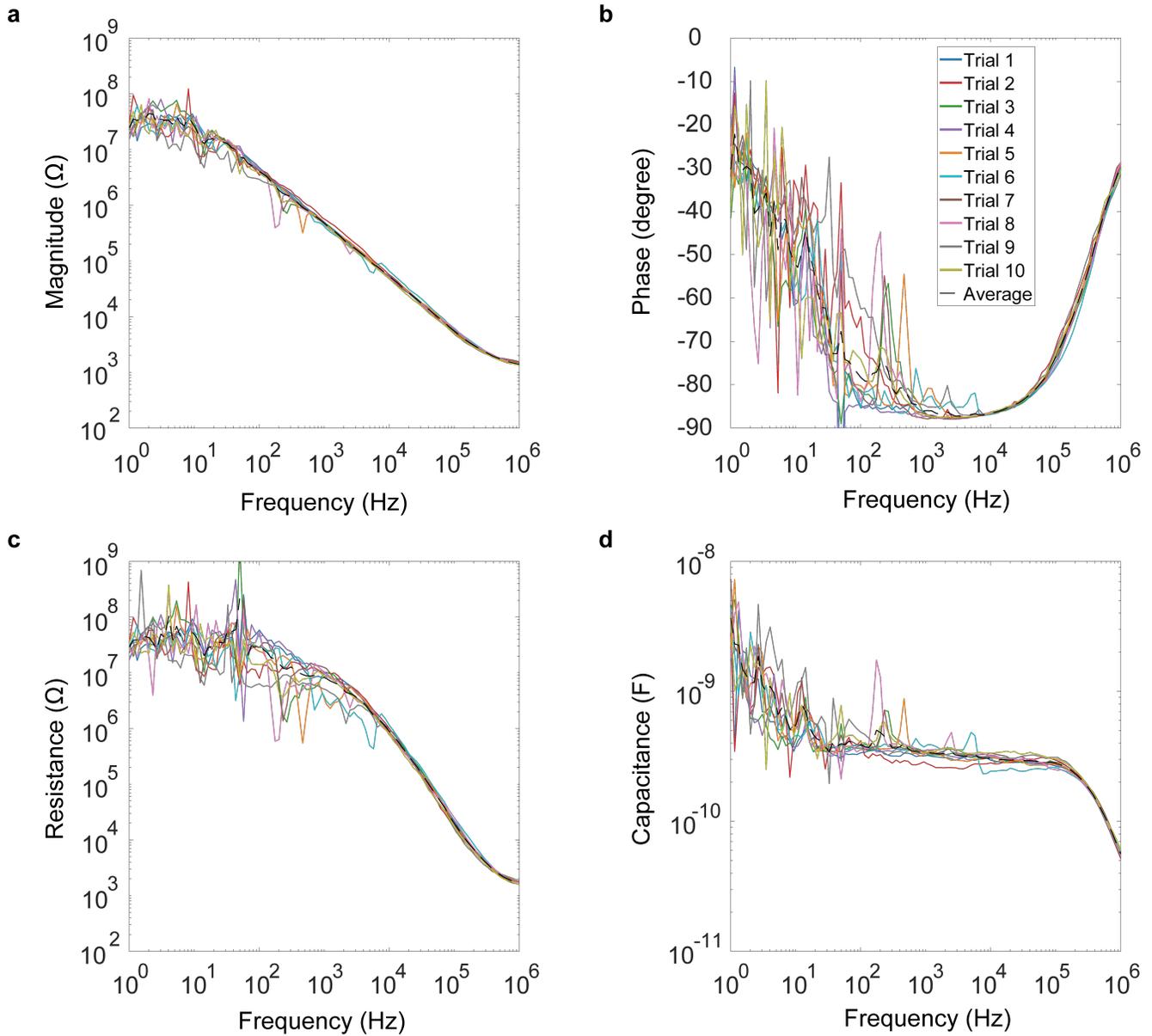

**Supplementary Figure S20.** Change in total impedance as a function of frequency for the finger sliding on the touchscreen under the nominal condition for day 3 (repeated ten times): a) magnitude, b) phase, c) resistance, and d) capacitance.

## SI. 6  Electrical Impedance Measurements for Finger Sliding on Touchscreen in Wet Condition

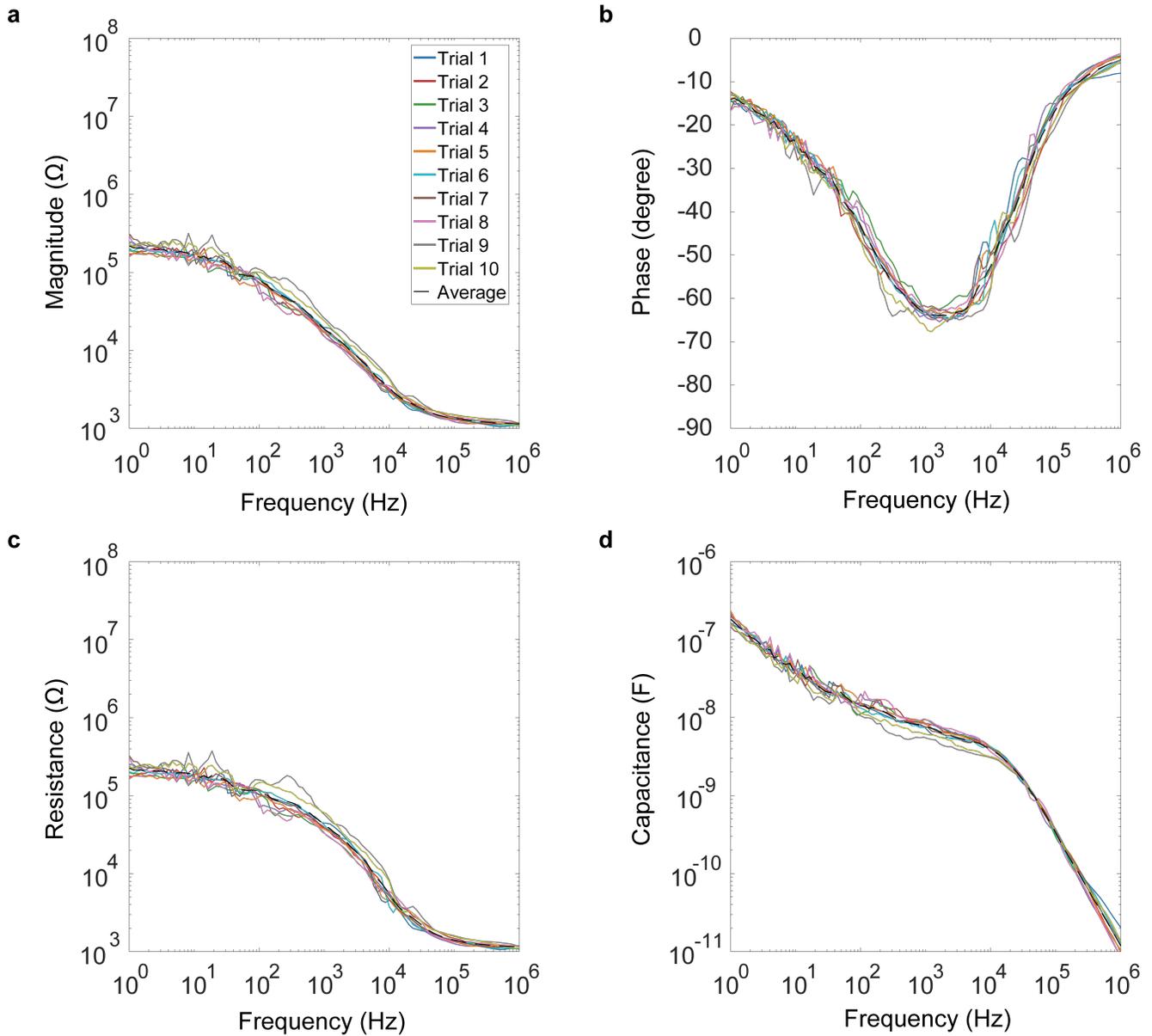

**Supplementary Figure S21.** Change in total impedance as a function of frequency for the finger sliding on the touchscreen under the wet condition for day 1 (repeated ten times): a) magnitude, b) phase, c) resistance, and d) capacitance.

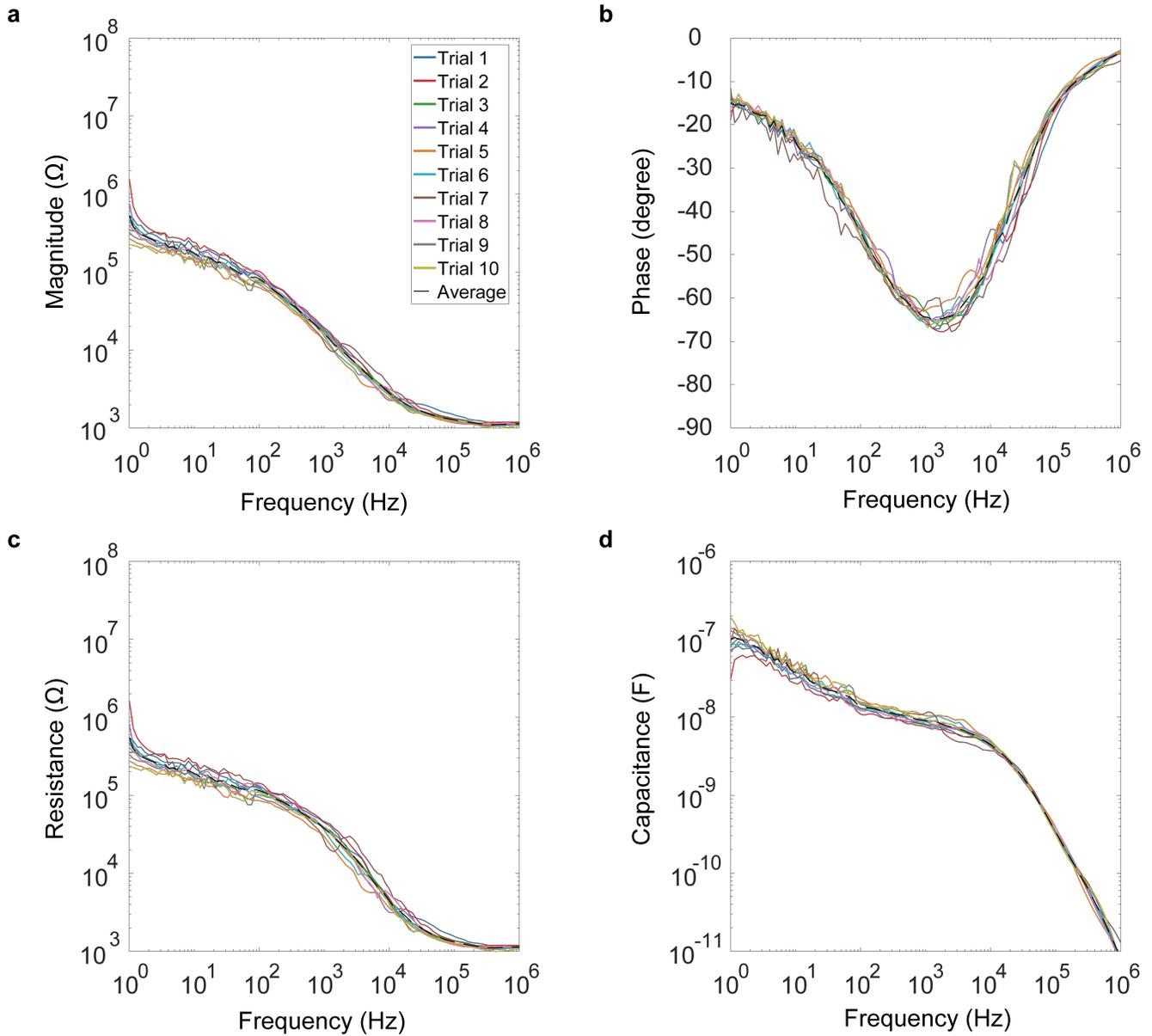

**Supplementary Figure S22.** Change in total impedance as a function of frequency for the finger sliding on the touchscreen under the wet condition for day 2 (repeated ten times): a) magnitude, b) phase, c) resistance, and d) capacitance.

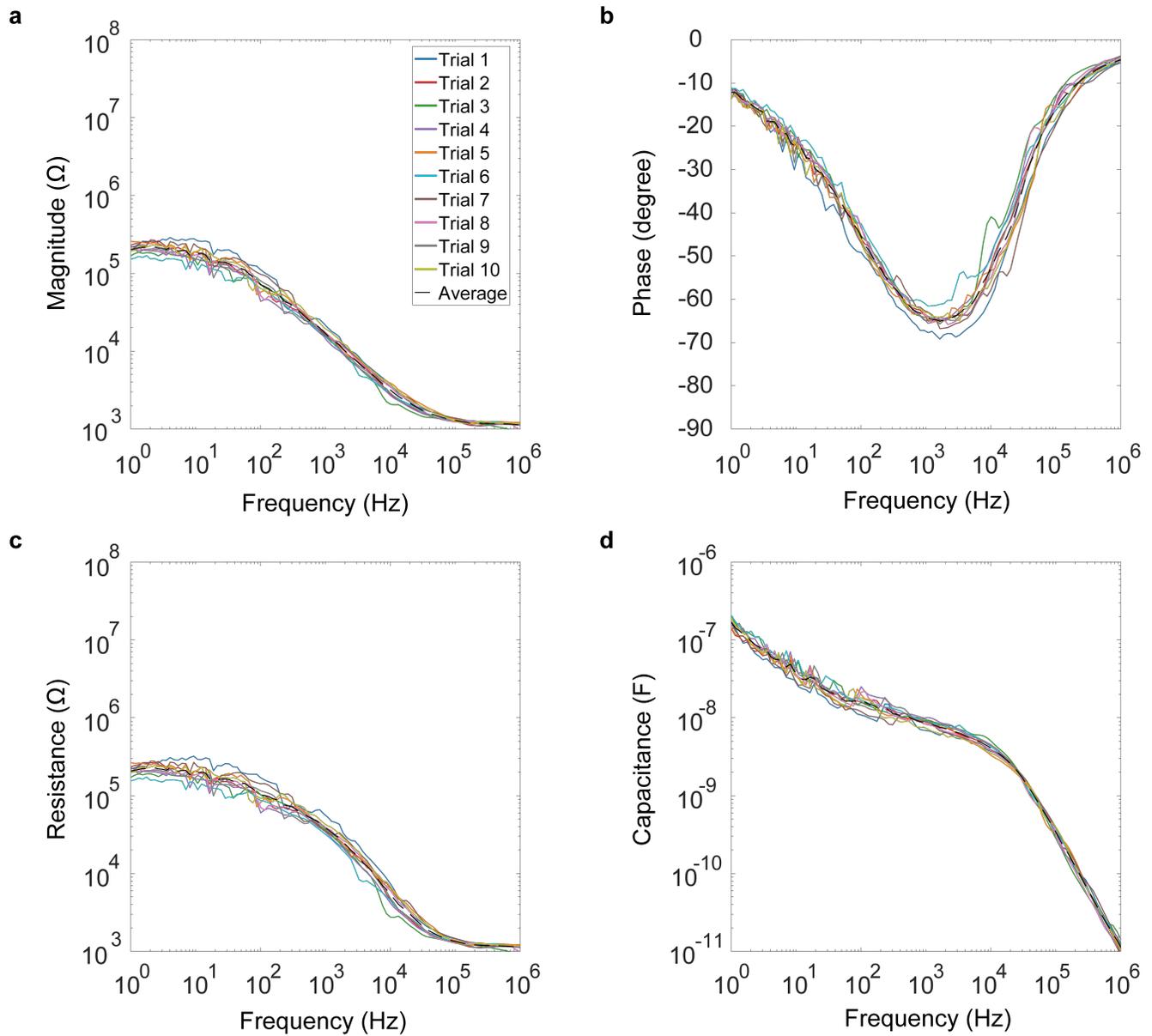

**Supplementary Figure S23.** Change in total impedance as a function of frequency for the finger sliding on the touchscreen under the wet condition for day 3 (repeated ten times): a) magnitude, b) phase, c) resistance, and d) capacitance.


\end{document}